\documentclass{aa}  
\usepackage{graphicx}
\usepackage{lscape}
\usepackage{natbib} 
\usepackage{siunitx}
\usepackage{url}
\usepackage{xcolor}
\usepackage{float}
\usepackage[varg]{txfonts}
\usepackage{multirow}
\usepackage[colorlinks=true,linkcolor=blue,citecolor=blue,urlcolor=blue]{hyperref}
\bibpunct{(}{)}{;}{a}{}{,} % citation like A&A style

   \def\HeI{\ion{He}{i}\,$\lambda$\,}
 \def\HeII{\ion{He}{ii}\,$\lambda$\,}

 \def\OIII{\ion{O}{iii}\,$\lambda$\,}

\newcommand{\msun}{\ifmmode M_{\odot} \else M$_{\odot}$\fi}
\newcommand{\rsun}{\ifmmode R_{\odot} \else R$_{\odot}$\fi}
\newcommand{\lsun}{\ifmmode L_{\odot} \else L$_{\odot}$\fi}
\newcommand{\zsun}{\ifmmode Z_{\odot} \else $Z_{\odot}$\fi}
\newcommand{\xsun}{\ifmmode X_{\odot} \else $X_{\odot}$\fi}
\newcommand{\msunpyr}{\ifmmode{\,M_{\odot}\,\mbox{yr}^{-1}} \else{ M$_{\odot}$/yr}\fi}
\newcommand{\velo}{\ifmmode\varv\else$\varv$\fi}
\newcommand{\vinf}{\ifmmode\velo_\infty\else$\velo_\infty$\fi}
\newcommand{\rgal}{\ifmmode \,R_{\mathrm{gal}} \else R$_{\mathrm{gal}}$\fi}

\begin{document}

\title{Determining stellar properties of massive stars in NGC{\,}346 in the SMC with a Bayesian statistic technique\thanks{Based on observations made with the NASA/ESA Hubble Space Telescope (HST), obtained from the data archive at the Space Telescope Science Institute. STScI is operated by the Association of Universities for Research in Astronomy, Inc. under NASA contract NAS 5-26555.}
}

   \author{M.\,J.~Rickard\inst{1,2}
        \and
          R. Hainich\inst{1}
          \and
          D. Pauli\inst{1}
          \and
          W.-R. Hamann\inst{1}
        \and
          L. M. Oskinova\inst{1}
          \and
          R. K. Prinja\inst{2}
          \and
          V. Ramachandran\inst{3}
          \and\\
          H. Todt\inst{1}
          \and
          E. C. Schösser\inst{3}
          \and
          A. A. C. Sander\inst{3}
          \and
          P. Zeidler\inst{4}
          }

   \institute{Institut f\"{u}r Physik und Astronomie, Universit\"{a}t Potsdam, Karl-Liebknecht-Str. 24/25, D-14476 Potsdam, Germany\\
              \email{matthew.rickard.18@ucl.ac.uk}
         \and
             Department of Physics and Astronomy, University College London, Gower Street, London WC1E 6BT, UK
        \and
            Zentrum für Astronomie der Universit\"{a}t Heidelberg, Astronomisches Rechen-Institut, M\"{o}nchhofstr. 12-14, 69120 Heidelberg, Germany
        \and 
        AURA for the European Space Agency (ESA), ESA Office, Space Telescope Science Institute, 3700 San Martin Drive, Baltimore, MD 21218, USA      
             }

   \date{Received 29 September 2024 / Accepted 17 October 2024}

%-------------------  Abstract --------------------

\abstract
%
% context
%
{NGC{\,}346 is a young cluster with numerous hot OB stars. It is part of the Small Magellanic Cloud (SMC), and has an average metallicity that is one-seventh of the Milky Way's. 
A detailed study of its stellar content provides a unique opportunity to understand the stellar and wind properties of massive stars in low-metallicity environments, and enables us to improve our understanding of star formation and stellar evolution.
} 
%
% aims heading (mandatory)
%
{The fundamental stellar parameters defining a star's spectral appearance are its effective surface temperature, surface gravity, and projected rotational velocity. Unfortunately, these parameters cannot be obtained independently from only  \ion{H}{}and \ion{He}{}spectral features as they are partially degenerate. With this work we aim to overcome this degeneracy by applying a newly developed Bayesian statistic technique that can fit these three parameters simultaneously.
}
%
% methods heading (mandatory)
%
{Multi-epoch optical spectra are used in combination with a Bayesian statistic technique to fit stellar properties based on a publicly available grid of synthetic spectra of stellar atmospheres.
The use of all of the multi-epoch observations simultaneously allows  the identification of binaries.

}
%
% results heading (mandatory)
%
{The stellar parameters for 34 OB stars within the core of NGC{\,}346 are derived and presented here. By the use of both \ion{He}{i} and \ion{He}{ii} lines, the partial degeneracy between the stellar parameters of effective surface temperature, surface gravity, and projected rotational velocity is overcome.
A lower limit to the binary fraction of the sample of stars is found to be at least 46\%.
}
%
% conclusions
%
{Based on comparisons with analysis conducted on an overlapping sample of stars within NGC{\,}346, the Bayesian statistic technique approach is shown to be a viable method to measure stellar parameters for hot massive stars in low-metallicity environments even when only low-resolution spectra are available.}

\keywords{Stars: mass-loss -- Methods: statistical -- Galaxies: star clusters: individual:NGC\,346} %Update these

\maketitle

%________________________________________________________________
\section{Introduction}
\label{sect:intro}

Massive stars have a large impact on the evolution of their environment, a disproportionately great effect given their relative scarcity. They drive ionisation feedback of interstellar material with their high ultraviolet (UV) flux \citep{HollenbachTielens1999, Matzner2002}. This high UV flux drives strong line-driven winds, depositing this shed wind material in the interstellar medium \citep{RogersPittard2013}. Many of the most massive stars are believed to end their lives as supernova explosions, which influence their environments by rapidly depositing huge amounts of energy and metal-enriched material \citep{RogersPittard2013}. Through these two mechanisms, massive stars are the main drivers of the chemical enrichment of the interstellar medium \citep{Burbidge+1957, Pignatari+2010, Thielemann+2011, Kasen+2017, Kajino+2019}.

One critical property that has a significant impact on the evolution of massive stars is metallicity; the chemical abundance of elements heavier than He. As these metals provide the transition lines used for accelerating the stellar wind \citep{LucySolomon1970, CAK1975}, the abundance of these metals has a significant impact on the proportion of the mass of a massive star lost through winds throughout its evolution \citep{Brott+2011, Georgy+2013, Pauli+2023}. 
Stellar winds can strongly impact a star's evolutionary pathway and its final stages, and defines the nature of the compact object that will be left behind \citep{Heger+2003}.
Despite this, the scaling of mass-loss rate with metallicity is poorly defined \citep{Abbott1982, Vink+2000, Vink+2001}. The need to resolve this has resulted in a significant quantity of work to study massive stars in low-metallicity environments \citep[e.g.][]{Bouret+2003, Martayan+2007, Ramirez-Agudelo+2013, Ramachandran+2018, Ramachandran+2019, Dufton+2019, Rickard+2022}.

The Small Magellanic Cloud (SMC)  is an irregular dwarf galaxy within the Local Group with a metallicity ${Z_\mathrm{SMC}=1/7\,\zsun}$ \citep{Trundle+2007, Hunter+2007}. It is close  enough to Earth for massive stars within it to be observed and analysed in great detail \citep[distance, $\mathrm{D}=61$\,kpc, distance modulus, $\mathrm{DM}=18.9\,\mathrm{dex}$;][]{Hilditch+2005}.
Due to a metal abundance that is  lower than Galactic massive stars, those in the SMC can maintain their mass for longer due to the  reduced mass loss through stellar winds. This not only impacts the stellar parameters of temperature, surface gravity, and luminosity, but also affects parameters such as rotational velocity, as less angular momentum is carried away, leading to faster rotating stars \citep{Martayan+2007, Ramirez-Agudelo+2013,  Dufton+2013, Ramachandran+2019}.

NGC{\,}346
% (Fig.~\ref{fig:hst}) 
is a stellar cluster within the \ion{H}{ii} region N\,66 that is  within the SMC. The OB star population of NGC{\,}346, a young cluster \citep[\,$\gtrsim 3\,\mathrm{Myr}$,][]{Rickard+2022, RickardPauli2023} in the SMC, has been a target of particular interest in the study of massive star winds at low metallicity, due to the abundance of mid-O-type  stars \citep{Massey+1989, Evans+2006, Dufton+2019, Rickard+2022}. Large samples of massive stars from within the Magellanic Clouds have been analysed based on their optical spectra \citep{Ramachandran+2018, Ramachandran+2019, Dufton+2019}.

The three stellar parameters, which have the largest impact on an OB star's spectral appearance are its temperature ($T_\ast$), surface gravity ($\log g$), and   projected rotational velocity ($\varv \sin i$). Unfortunately, these three stellar parameters are affectbed by partial degeneracy when measured based on \ion{H}{}and \ion{He}{}spectral lines. This means they cannot be fitted independently as
each of these stellar properties has a measurable impact on the depth and width of photospheric \ion{H}{}and \ion{He}{}spectral lines. The parameters $T_\ast$ and $\log g$ affect the depth, and  $\log g$ and $\varv \sin i$  both impact the line broadening. To overcome this partial degeneracy, the usual method is to measure $\varv \sin i$  independently of  the \ion{He}{}~lines by fitting the line broadening of metal lines.
 The best-fit $T_\ast$ and $\log g$ is then found by comparing observations to a grid of synthetic spectra generated from stellar atmosphere model codes such as the Potsdam Wolf-Rayet code \citep[PoWR, ][]{Grafener+2002, HamannGrafener2004, Oskinova+2011, Hainich+2014, Hainich+2015, Shenar+2015, Sander+2015, Hainich+2015}, \textsc{cmfgen} \cite{Hillier1987, HillierMiller1998, HillierLanz2001, Hillier2012}, and \textsc{tlusty} \citep{Hubeny1988, HubenyLanz1995, Hubeny+1998, LanzHubeny2007}, among  others.

The resultant fit for the stellar parameters for each target in a large sample forms the starting approximation. The next step would then be to carry out a significant manual adjustment to fit a suitable set of model parameters. 
As a result, studying populations of massive stars represents a significant time investment. Efforts have been made to automate this time consuming process, for example HiLineThere \citep{Rubke+2023}, which automates the line selection process, the measurement of $\varv \sin i$, and the selection of the most appropriate model from a grid of synthetic spectra based on a $\chi^2$ technique.

The measurement of $\varv \sin i$ is complicated by additional mechanisms that broaden the profiles of absorption and emission lines within the star's observed spectrum. Rotation broadening dominates the broadening effects, with rotation velocities being of the order of 10s or 100s of $\mathrm{km\,s}^{-1}$ up to the critical rotation velocity. The additional broadening mechanism have magnitudes typically of the order of a few $\mathrm{km\,s}^{-1}$ \citep{Aerts+2009}. The term adopted for these additional broadening mechanisms is macroturbulence, even though it is  likely that it  is not large-scale turbulent motion \citep{Simon-DiazHerror2014}. With high-resolution spectra it is possible to measure the rotational broadening separately from the combined broadening effects of both rotation and macroturbulence. This has been done for massive stars in NGC\,346, and the difference between the two measurements was found to be $\sim{\,}11{\,}\mathrm{km\,s}^{-1}$ \citep{Dufton+2019}. When comparing results between numerous sources, it is important to consider if the measurement of $\varv \sin i$ given excludes macroturbulence or includes macroturbulence, and thus it must be considered a likely   small overestimation.

In this work, we present a novel approach to automating the measurement of the stellar parameters of $T_\ast$, $\log g$, and $\varv \sin i$  of massive stars. This uses the Bayesian statistic technique of Markov chain Monte Carlo (MCMC), while also accounting for the radial velocity (RV) of each object in each observation. The aim is to develop a tool applicable for low-resolution spectra where narrow metal lines are not visible, overcoming the partial degeneracy between these stellar parameters.
We employ multi-epoch observations to mitigate the inaccuracies introduced by noise across different observations and to measure the RV movement between epochs and identify binary candidates.

The paper is organised as follows. Section~\ref{sect:obs} describes the observations before setting out the Bayesian statistic technique in Section~\ref{sec:analysis}. The results are described in Section~\ref{sec:results}, with the discussion of these results following in Section~\ref{sec:discussion}. Our conclusions are presented in Section~\ref{sec:conclusions}. The observation list and full results table are available in Appendix~\ref{app:additional_tables}. Further supporting plots are available on Zenodo,\footnote{\url{10.5281/zenodo.13991997}} including the results of individual stars and observations of miscellaneous targets.

%________________________________________________________________
\section{Observations and data reduction}
\label{sect:obs}

NGC{\,}346 was observed with the Multi Unit Spectroscopic Explorer (MUSE), an integrated field unit spectrograph (IFU), on the European Southern Observatory (ESO) Very Large Telescope (VLT) between August 11 and 22, 2016 (ESO programme 098.D-0211(A), PI W.-R. Hamann). 
While previous studies   employed this data \citep{Zeidler+2022}, for this work we included new extractions of spectra for each target. Ultraviolet (UV) spectra previously presented in \citet{Rickard+2022} were also utilised.

\subsection{MUSE observations}
\label{sec:MUSE_obs}

The MUSE observations were taken in wide-field mode (WFM). This field of view (FOV, 1\arcmin) covers the central part of the giant H\,II region N\,66 that is powered by the massive star cluster NGC{\,}346.
MUSE has a spectral resolution of $R \sim 2000{-}4000$  and a wavelength range of ${\sim}4800{-}9300\,\AA$ \citep{MUSE2010}. Altogether 11 observing blocks (OBs), consisting of eight science exposures of 315\,s each, were obtained. 
The observing conditions for the individual OBs are listed in Table~\ref{table:obs_cond} in Appendix~\ref{app:additional_tables}, showing how the seeing varied between 1.17\arcsec and 1.92\arcsec.
The observations were carried out without adaptive optics, which was not yet available in 2016.

NGC\,346 has strong nebular emission, including [\ion{O}{iii}] $\lambda$ 5007, \HeI{7065,} and $\mathrm{H}\alpha$.
The inhomogeneity of the nebula makes it difficult to effectively remove the nebular lines from the stellar spectra. A background subtraction was tested using the subtraction function provided by \textsc{PampelMuse}, but nebular lines were often still present, with under- or over-subtractions affecting these lines. Thus, the effort was abandoned and instead the presence of nebular features were considered when selecting which photospheric lines to fit to.
For example, only the wings of H$\beta$ and H$\alpha$ lines were considered due to nebular H features (see Sect.~\ref{sec:line_segment_selection} for the full description of the lines selected).

Using the \cite{Sabbi+2007} catalogue as the input to the pipeline, a non-flux calibrated spectrum for each target position from each OB was produced, regardless of quality of the output. The signal-to-noise ratio (S/N) for each spectrum is calculated as the mean S/N for a small selection of continuum regions based on the MUSE pipeline, free from nebular contributions. A S/N cutoff is employed, keeping only spectra with $\mathrm{S/N} > 50$. A total of 226 targets are found to have at least one MUSE observation extracted that fulfils this S/N criteria.

\begin{figure}
\centering
\includegraphics[width=\hsize]{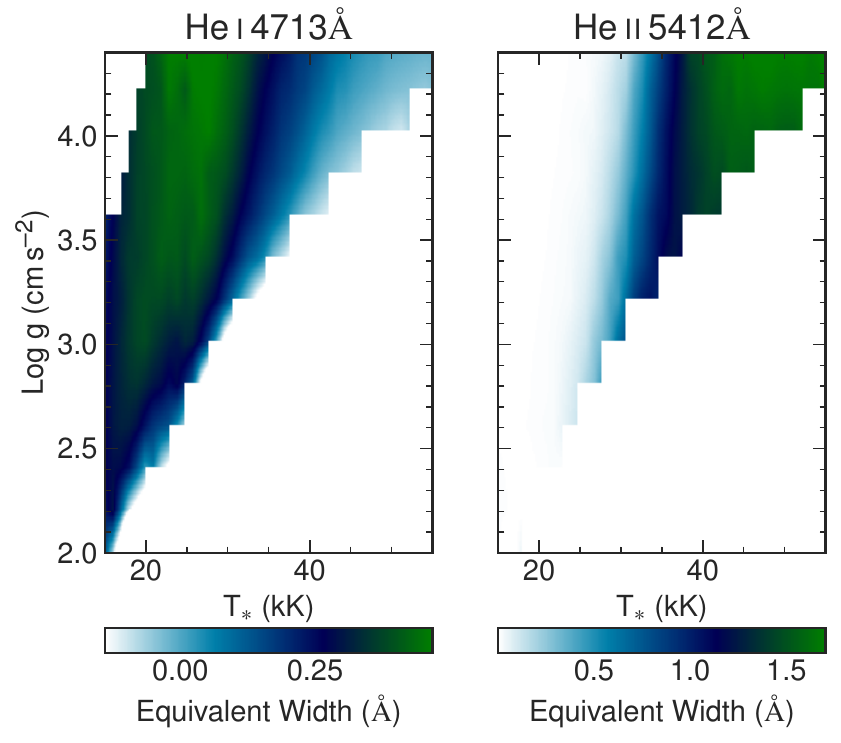}
 % \includegraphics[width=0.495\hsize]{4713_EW_grid.pdf}
 % \includegraphics[width=0.495\hsize]{5412_EW_grid.pdf}
 % \includegraphics[width=\hsize]{4713_EW_grid.pdf}
 % \includegraphics[width=\hsize]{5412_EW_grid.pdf}
% \sidecaption
 % \includegraphics[width=6cm]{4713_EW_grid.pdf}
 % \includegraphics[width=6cm]{5412_EW_grid.pdf}
    \caption{
                Equivalent widths of \HeI{4713} (left) and \HeII{5412} (right), bilinearly interpolated for the whole PoWR grid range. 
        }
    \label{fig:EW_interpolated}
\end{figure}
%-------------------

Initially, the number of spectra to consider numbered in the thousands. Therefore, for the normalisation, an automated process was employed. This enabled a first attempt at the analysis using the MCMC method described in Section~\ref{sec:analysis}. This first pass allowed the characteristics of the process to be considered, including limiting factors such as the requirement for some \ion{He}{ii} lines to be present in the observations. This allowed the targets where the Bayesian statistic technique would be successfully  identified. The number of spectra when limiting the selection to these targets alone was $~\sim 300$.
% where the method could be applied identified based on these requirements, the number of spectra was reduced to $~\sim 300$.
These spectra were then normalised again with a visual check and adjustment of the wavelengths of the normalisation points to improve the quality of the data used in the final pass of the Bayesian statistic technique.

The process of normalising a spectrum introduces an error beyond the calculated S/N. The error on the normalised flux of each spectrum is calculated from the standard deviation of four continuum regions absent of lines, each $50\AA$ wide, centred on 4500\AA, 4805\AA, 5450\AA,~and 6485\AA. These regions are selected as broad regions devoid of lines, but located near the lines employed later for fitting. This value is used within the likelihood function of our fitting tool to weight the observations based on the flux error (Sect.~\ref{sec:Bayesian_process}).

\subsection{Ultraviolet observations}
\label{sec:UVobs}

The brightest objects in the core 1\arcmin~of NGC\,346 were previously   observed in UV with the Hubble Space Telescope  (HST; GO 15112, PI L. Oskinova \& GO 8629, PI F. Bruhweiler). These observations were presented in \citet{Rickard+2022}
These spectra are included in this work to allow   the measurement of luminosity and extinction. These HST observations are single-source extractions from the long-slit G140L on STIS. The extracted UV spectra cover an approximate range $\lambda \, 1150 - 1700 \, \AA$, with some variation depending on the stars' offset from the centre of the observing slit. The spectra have the resolution $\lambda / \Delta \lambda \sim 2400$. The extraction of this data is described in more detail in \citet{Rickard+2022}.

\section{Data analysis}
\label{sec:analysis}

Each target within the sample was analysed with a Bayesian statistic technique using the Python package for MCMC methods \textsc{emcee} \citep{MCMChammer}. In the language of the Bayes theorem, the data (the MUSE observations) are compared to a hypothesis (a synthetic spectrum created from a set of stellar parameters, see Sect.~\ref{sec:synth_spec}). In this way, the best-fitting stellar parameters, the ones that create the best-fitting synthetic spectrum, can be identified. This method was designed here to rely entirely on H and He lines within the spectra of the target stars. 
Due to the low metallicity of our targets and the high temperature, very few metal lines are detectable in the MUSE wavelength range. Even when a metal line is present in the spectrum of some target stars (such as \OIII{5591}), it is very weak and not present for the full sample of targets. Thus, for consistency, the method was designed to not require any metal lines.

Each observed spectrum was included independently, each shifted by the RV shift for each epoch. This created an additional free parameter, the RV shift, per epoch.

\subsection{The hypothesis: Synthetic spectrum generation}
\label{sec:synth_spec}

The hypothesis is a synthetic normalised spectrum generated based on $T_\ast$ and $\log g$ and then convolved to simulate $\varv \sin i$  and convolved to account for the instrument's observation profile. The interpolated normalised synthetic spectrum is based on a grid of synthetic spectra generated by the PoWR code model, presented in publicly available grids.\footnote{\url{www.astro.physik.uni-potsdam.de/PoWR/}} The grids of synthetic OB-type star spectra have a step of 1000 K in $T_\ast$ and 0.2 dex in $\log g$. Within the PoWR grid, $T_\ast$ is defined as the temperature at the Rosseland continuum optical depth.
As the sample is within the SMC, the `SMC OB Vd3' grid (Pauli et al. in prep.) was utilised. In this grid, the wind mass-loss rates employed in the model calculations use the recipe from \citet{Vink+2000, Vink+2001} divided by a factor of three. From previous studies of stars in NGC{\,}346 \citep{Rickard+2022}, this grid was selected as the most suitable of the available options as no obvious wind effects are seen in the optical to near-infrared spectra of these massive stars.

An interpolated synthetic normalised spectrum can be created for any $T_\ast$ and $\log g$ within the limits of the PoWR SMC grid by bi-linearly interpolating between the four synthetic spectra flanking the required $T_\ast$ and $\log g$ values. Figure~\ref{fig:EW_interpolated} shows the equivalent widths (EWs) of two He lines for the full parameter space, interpolated between grid points.

The synthetic spectrum is sliced as per the observed spectra (See Sect.~\ref{sec:line_segment_selection}).  Each segment is convolved with a half ellipse with the equivalent width  associated with a value of $\varv \sin i$ to simulate the effect of rotation and other broadening mechanisms, such as macroturbulence.
The rotational broadening velocity will be of the order of tens or hundreds of $\mathrm{km\,s}^{-1}$ up to the critical rotation velocity, while the macroturbulent broadening is of the order of $10 \, \mathrm{km\,s}^{-1}$ \citep{Aerts+2009, Dufton+2019}. It is therefore important to be aware that the values of $\varv \sin i$ returned from this method will be a slight overestimation of approximately $10 \, \mathrm{km\,s}^{-1}$. This is important to consider when comparing results to other sources.

Finally, the instrument profile is simulated by convolving the synthetic spectral line segment with a Gaussian of FWHM appropriate for the instrument profile for MUSE at the wavelength of that line segment. The MUSE documentation shows the instrument profile changes linearly with wavelength up to $\lambda ~\sim 7000 \AA$ and non-linearly beyond that (see Figure 18 in the MUSE User Manual\footnote{\url{https://www.eso.org/sci/facilities/paranal/instruments/muse/doc/ESO-261650_MUSE_User_Manual.pdf}}). The adopted FWHM for each line segment is shown in Table~\ref{tab:lines_segments}.

\subsection{Line segment selection}
\label{sec:line_segment_selection}

A critical choice for the process of comparing the observations to a synthetic spectrum is the choice of wavelength ranges used for the comparison. The line segments used for fitting are detailed in Table~\ref{tab:lines_segments}. These were carefully selected to include a selection of \ion{He}{i} and \ion{He}{ii} lines, as well as $\mathrm{H}\beta$ and $\mathrm{H}\alpha$ lines. The line boundaries were chosen to exclude nebular emission features. To do this, the $\mathrm{H}\beta$ line was split into two segments (to exclude nebular $\mathrm{H}\beta$ emission) and $\mathrm{H}\alpha$ was split into four segments (including \HeII{6529.5} which overlaps with the blue wing of $\mathrm{H}\alpha$).

The \HeII{4686} line was excluded as it can be seen that on the hotter stars there is some back filling due to stronger stellar winds than included in the model. In addition, some single \ion{He}{i} lines such as \HeI{4922} are not well modelled by codes such as PoWR and \textsc{cmfgen} \citep{Najarro+2006} and were therefore not selected.

\begin{table} 
\footnotesize
    \centering 
    \caption{Line segments for fitting.} 
    \begin{tabular}{cccc} \hline \hline \rule{0cm}{2.2ex} 
    Central & Window & Instrument & Line \\
    Wavelength & Width & Profile (FWHM)  & \\ 
    $\AA$ & $\AA$ & $\AA$ \\ 
    \hline \rule{0cm}{2.4ex} 
    4845.0 & 24.0 & 2.84 & $\mathrm{H}\beta$ blue \\
    4877.0 & 24.0 & 2.84 & $\mathrm{H}\beta$ red \\
    4713.2 & 6.0 & 2.87 & \ion{He}{i} \\
    5016.7\textsuperscript{$\dagger$} & 5.0 & 2.81 & \ion{He}{i} \\
    5412.0 & 19.0 & 2.74  & \ion{He}{ii} \\
    6529.5 & 14.5 & 2.60  & \ion{He}{ii} \\
    6554.0 & 3.0 & 2.60 & $\mathrm{H}\alpha$ blue \\
    6574.0 & 6.0 & 2.60 & $\mathrm{H}\alpha$ red \\
    6592.9 & 5.9 & 2.60 & $\mathrm{H}\alpha$ red \\
    \hline \rule{0cm}{2.4ex} 
    \end{tabular} 
    \tablefoot{\tablefoottext{$\dagger$}{Offset to exclude blueward nebular contamination.}}
    \label{tab:lines_segments} 
    \rule{0cm}{2.8ex}
\end{table}

\subsection{Determining stellar parameters using a Bayesian statistic technique}
\label{sec:Bayesian_process}

The Bayesian statistic technique measures the quality of the fit of the data, H and He line segments from up to 11 observations, against the {hypothesis, the synthetic spectrum generated based on a set of stellar parameters. In doing so, the process  identifies the three best-fitting  stellar parameters that generate the synthetic spectrum ($T_\ast$,  $\log g$, and $\varv \sin i$) and the best-fitting RV for each observation. This means there may be as many as 14 free parameters (when a target has 11 observations with sufficient S/N) or as few as 4 (if there is only one usable observation). The data is measured against the hypothesis by  way of a least mean squared likelihood function (Eq.~\ref{eq:lnlike}):

\begin{equation}
    \label{eq:lnlike}
    \ln \, \mathbb{P}\,(F_{\lambda, \mathrm{synth}}|F_{\lambda, \mathrm{obs}}, \sigma F_{\lambda, \mathrm{obs}}) = - \frac{1}{2} \sum\limits_{\lambda}
    \left(\frac{F_{\lambda, \mathrm{obs}} - F_{\lambda, \mathrm{synth}}}{\sigma F_{\mathrm{obs}}}\right)^2
.\end{equation}

A flat prior probabilty is assumed for all the fitted parameters, with $T_\ast$ and $\log g$ values limited to the range of the PoWR grid. $\varv \sin i$  is limited to ${10\,\mathrm{km \, s}^{-1}< \varv\,\sin\,i <600\,\mathrm{km \, s}^{-1}}$ and the RV of each observation is limited to ${90\,\mathrm{km \, s}^{-1}<\mathrm{RV}<240\,\mathrm{km \, s}^{-1}}$, a wide range centred on $165\,\mathrm{km \, s}^{-1}$, the mean RV value for OB-star members of NGC{\,}346 \citep{Zeidler+2022}.

The parameter space is seeded with 250 `walkers' or chains. Each chain starts with a set of initial values for the stellar parameters and for the RV of each observation. This set of initial values is randomly distributed across the space described by the flat prior. The MCMC process tests these parameters against the observations using the likelihood function. The automated chain process then tests another set of parameters, adjusted from the first by a random offset. The new likelihood is calculated and compared to the previous, to evaluate whether the new parameters are a better fit or a worse fit than the previous parameter set. This then informs the next set of parameters to be tried and assessed with the likelihood function, and again compared to the previous likelihood. In this way, these chains `walk' towards the more likely solution, as defined by the likelihood function.

For an MCMC process, the auto-correlation time ($\tau$) is a measure of how many steps a walker will need to take until the initial position has no bearing on the parameters it is testing, when it is said to have `forgotten' its starting position. The value of $\tau$ can be calculated during the process of running an analysis within \textsc{emcee}. The auto-correlation time is used here to dynamically set the number of iterations needed to be run and to estimate when additional iterations will not improve the final result, meaning the process can be halted and said to have converged. Every 100 iterations, $\tau$ is calculated and the result is considered converged if either   the sampler has run for $> 20 \cdot \tau$ for each fit parameters or   $\tau$ has changed by less than 1\% for each fit parameter over the last 100 iterations compared to the previous 100 iterations. Once complete, the first $5\cdot\Bar{\tau}$ steps are discarded to allow for burn-in, to ensure that there is no impact on the final result from the starting positions, and the sample is thinned by $\Bar\tau/\/2$, for reasons of computational speed.

% ---------------------------------------------------
\begin{figure}
        \centering
        \includegraphics[width=\hsize]{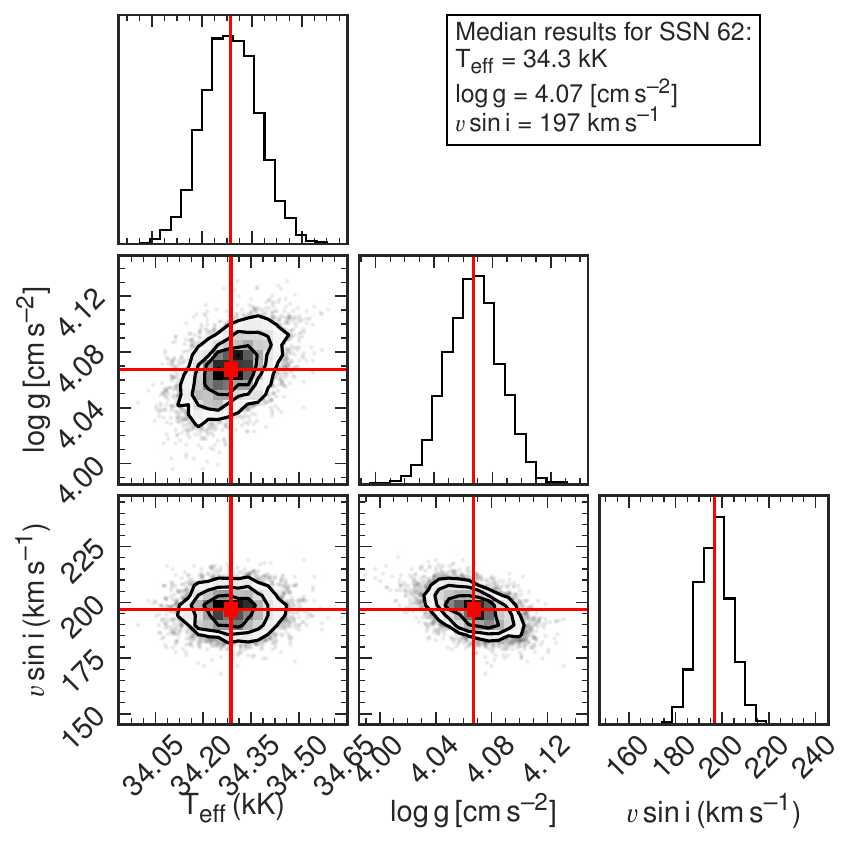}
        \caption{
                Corner plot showing the PDF of the sampler results of the three stellar parameters for SSN~62. This is an example of a well-converged single-star fit.
        }
        \label{fig:SSN62_stellar_param_corner_plot}
\end{figure}
%--------------------

%---------------------------------------------------------------
\begin{figure}
        \centering
        \includegraphics[width=\hsize]{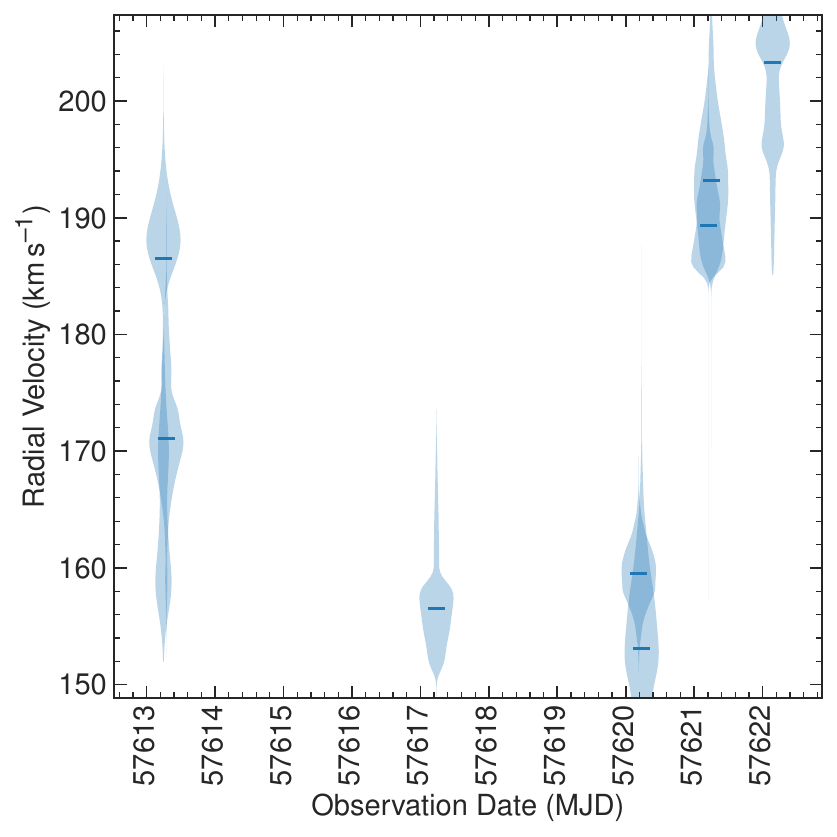}
        \caption{
                Violin plot of sampler RV returns for each MUSE observation of SSN~62. This is an example of an SB1 as at least two median RVs have a separation of $>10 \mathrm{km \, s}^{-1}$ and the RV differences in two epochs is significant with a separation $>4\sigma$.
        }
        \label{fig:SSN62_vrads_violin_plot}
\end{figure}
%---------------------------------------------------------------

%---------------------------------------------------------------
\begin{figure*}
        \centering
        \includegraphics[width=\hsize]{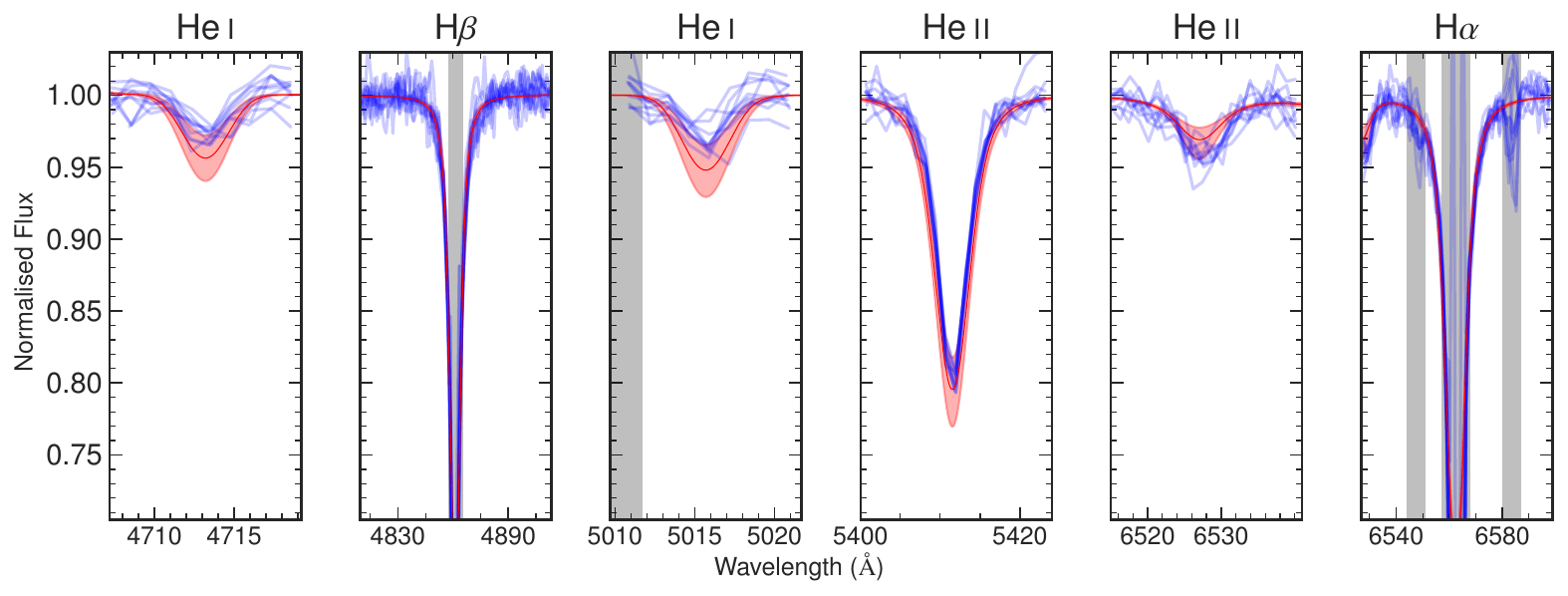}
        \caption{
                Synthetic spectrum created from the median sampler results for SSN~22 (\emph{red}), with the normalised MUSE observations (\emph{blue}), each observation shifted by the median RV sampler result for that observation. The red shading above and below the synthetic spectrum indicates the limit of all error bounds combined (as determined in the method described in Section~\ref{sec:error_estimation}), also including the largest error for RV of any observations. The shaded wavelength regions are not within the selected line regions detailed in Table~\ref{tab:lines_segments} and so are not considered in the fitting process.
        }
        \label{fig:SSN22_lines_comparisons_plot}
\end{figure*}
%--------------------

%---------------------------------------------------------------
\begin{figure}[tbp]
        \centering
        \includegraphics[trim=.1cm .2cm .3cm .5cm,clip,width=0.95\hsize]{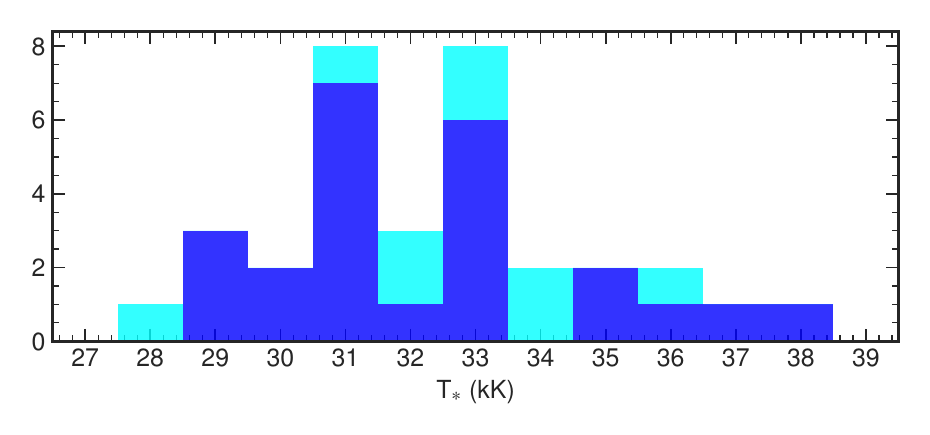}
        \includegraphics[trim=.1cm .2cm .3cm .5cm,clip,width=0.95\hsize]{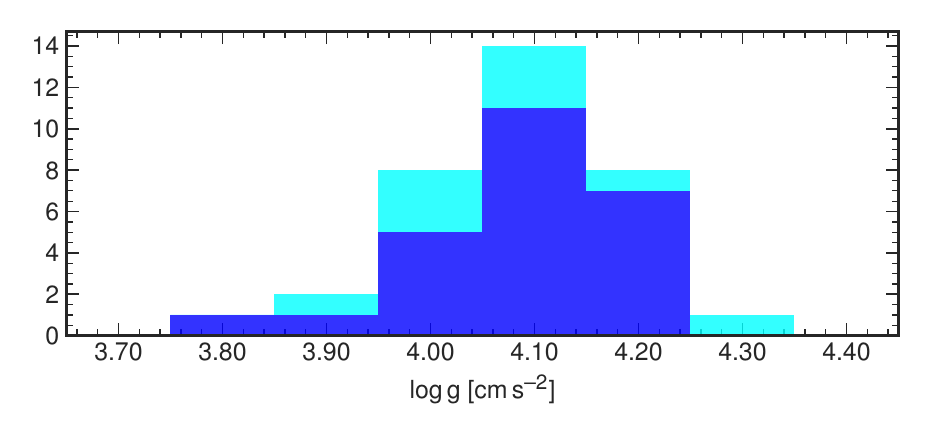}
        \includegraphics[trim=.1cm .2cm .3cm .5cm,clip,width=0.95\hsize]{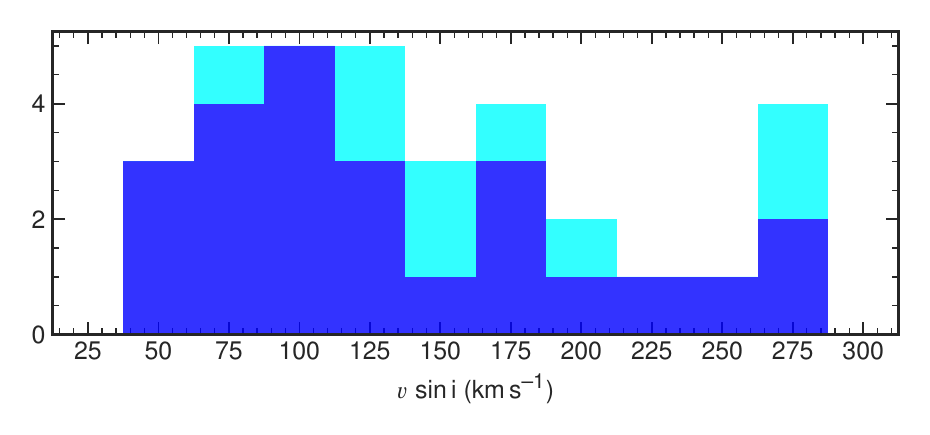}
        \includegraphics[trim=.1cm .2cm .3cm .5cm,clip,width=0.95\hsize]{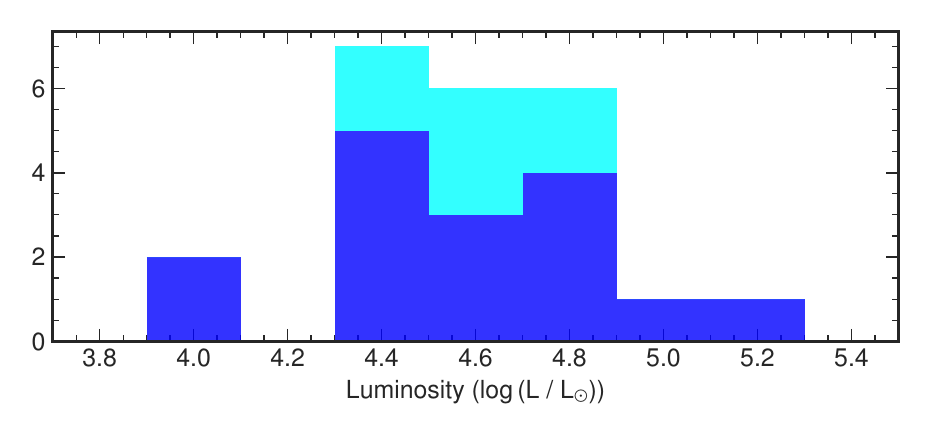}
        \includegraphics[trim=.1cm .2cm .3cm .5cm,clip,width=0.95\hsize]{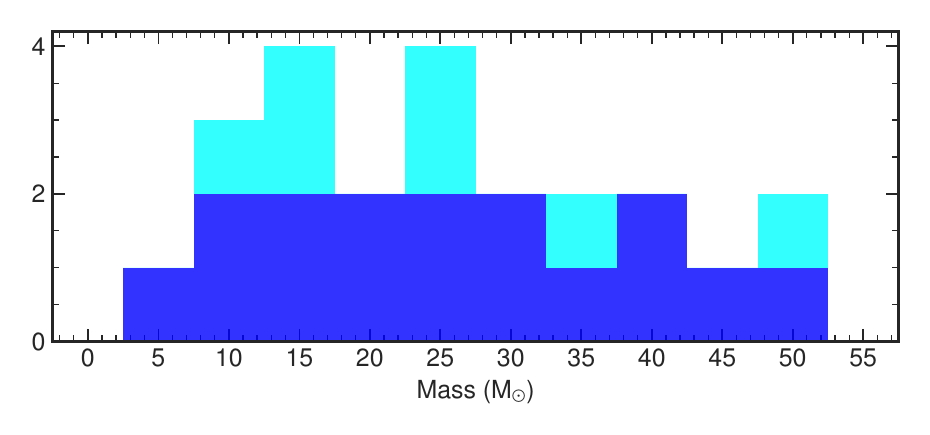}
        \caption{Distribution of the stellar parameters of the sample shown as stacked histograms of the sample. \emph{Blue}: Single stars, no evidence of multiplicity found. \emph{Cyan}: SB1s based on RVs.}
        \label{fig:stellar_params_hist}
\end{figure}
%---------------------------------------------------------------
%------------

\subsection{Convergence checks}
\label{sec:convergence}
As these results are obtained from an automated procedure, the sampler output for each target is closely inspected in the following ways to ensure confidence in the results for each individual target: a corner plot of the three stellar parameters (e.g. Figure~\ref{fig:SSN62_stellar_param_corner_plot}), a violin plot of the RV parameter sampler (e.g. Figure~\ref{fig:SSN62_vrads_violin_plot}), and a plot showing the observations compared the best-fitting synthetic spectrum. This is created by interpolating from the median $T_\ast$ and $\log g$ sampler results and broadened the synthetic spectrum generated by the median $\varv \sin i$  and convolving by the MUSE instrument profile. This is then plotted against the normalised MUSE observations, each shifted by the corresponding median RV (e.g. Figure~\ref{fig:SSN22_lines_comparisons_plot}).
This inspection can result in a seemingly converged result being rejected even when the 
$\tau$-convergence criteria has been reached.

A result may be rejected if the corner plot of the three stellar parameters show that the result is on the edge of the grid. This indicates that the degeneracy between the three parameters has not been resolved. An inspection of these cases shows that this occurs when there is insufficient \ion{He}{ii} line strength. This results in the inability to fit temperature.

\subsection{Error estimation}
\label{sec:error_estimation}

Typically from an MCMC process, the statistical upper and lower errors are taken from the quartiles (typically the 16{th} and 84{th}) of the sampler results. In this case, the statistical errors of this method would be of the order of $10\,\mathrm{K}$ for $T_\mathrm{eff}$, less than 0.01 dex $[\mathrm{cm\,s^{-2}}]$ for $\log g$, and only a few $\mathrm{km\,s}^{-1}$ for $\varv \sin i$. It is clearly not suitable to consider these minute statistical errors as the errors on the fit of the physical parameters of the targets.

To provide an estimation of the errors on the stellar parameters, we consider the variations in the found best-fit parameters that may occur from fitting individual observations. While the method described in Section~\ref{sec:Bayesian_process} considers all observations simultaneously to find the model that best fits all observations, weighted by the S/N of each, it is also possible to repeat the process for each observation individually. The result is then a separate sampler return for each observation, and, after discarding the burn-in, the median of the sampler for each stellar parameter can be taken to find the best-fit value from that observation alone. The standard deviation of the set of each stellar parameter provides a stronger estimate of the error, as it shows the variation in the best-fit parameter that may have been found if only considering one observation at a time. This method for each target in estimated errors on the order of $~200 - 1200 \,\mathrm{K}$ for $T_\mathrm{eff}$, 0.05 - 0.20 dex $[\mathrm{cm\,s^{-2}}]$ for $\log g$ and 10 - 80 $\mathrm{km\,s}^{-1}$ for $\varv \sin i$. These error estimation for each target are included in Table~\ref{tab:NGC346_MUSE_results} in Appendix~\ref{app:additional_tables}.

%________________________________________________________________

\subsection{Luminosity and extinction}
\label{sec:lum_fit}

Luminosity and extinction are constrained independently from the Bayesian statistic technique by using the UV spectral observations \citep{Rickard+2022}, along with HST F225W, F555W, and F814W photometry \citep{Sabbi+2007, Rickard+2022}.
The PoWR grid model with the closest parameters to the fit found for each target is selected, and the synthetic emergent spectral energy distribution (SED) for that model is compared to the observed HST UV spectra and HST photometry.

The adopted reddening value strongly influences the slope of the UV spectrum for each object, and thus the luminosity and extinction must be found simultaneously. Extinction is applied to the synthetic SED following the same method as in \citet{Rickard+2022}, which incorporates two elements. The first is a foreground Galactic extinction component, constant for all targets at $E(B{-}V)=0.06$\,mag, applied with the extinction law from \citet{Seaton1979}. The second is a local extinction 
specific for each star. This is applied using the extinction law for the SMC from \citet{Howarth1983} with $R_V = 2.7$ \citep{Bouchet+1985}. Both luminosity and $E(B{-}V)$ are adjusted simultaneously until the synthetic SED matches the observations. Where there is disagreement between the best extinction value to match the photometry and the HST UV SED, a preference is given to the best value for the UV SED observations.
Where there is no UV spectrum for the target, no luminosity or extinction is given.

\subsection{Spectral typing}

Given the lack of metal lines in the low-resolution MUSE spectra, a classical spectral classification that uses line ratios of metal lines is not possible. To approximate the spectral type here, we use the approximation for SMC stars reported by \citet{Ramachandran+2019}. Without any obvious giants, the relation for dwarfs and subdwarfs between O3 and B1 is used ($T_\ast \mathrm{[kK]} = 56.60 - 2.74 \times \mathrm{ST}$, where ST is the spectral type given as a number beginning with O3 = 3 and ending with B1 = 11).

\subsection{Binary candidate identification}
\label{sec:sb1}

Single-line spectroscopy binary (SB1) candidates are determined from the parameters for each observation from the return of the MCMC sampler. The median RV from the sampler return is the RV result for each observation with the standard deviation as the error. The criteria to identify targets with significant RV variation to consider SB1 candidates is taken from \citet{Dufton+2019} and \citet{Sana+2013}. From \citet{Dufton+2019}, we adopt the minimum variation to be considered an SB1 candidate to be at least $10 \,\mathrm{km \, s}^{-1}$ between two median RVs. As the targets all appear to be dwarfs, such a low number is suitable as they do not have strong intrinsic variability. From \citet{Sana+2013} we adopt the significance check used, where the RV differences between two epochs are significant if the separation is at least $4\sigma$.

The violin plot (e.g. Fig.~\ref{fig:SSN62_vrads_violin_plot}) of the sampler RV returns for each target is inspected. This is required as some of the observations within this study follow on from one another.
The RVs of these sequential observations usually agree with each other, but not always, for example when one of them has a low S/N (see Sect.~\ref{sect:obs} for the discussion about observing conditions). Some targets show erroneous median RV results between the results for two consecutive observations. If an erroneous median RV has caused a target to be classified as an SB1 candidate, this status is then revoked based on this inspection.

In addition, SB2 and even SB3 candidates can be identified by inspection of the morphology of the absorption lines. Multiple absorption peaks may be observed in one epoch, while others may show the peaks overlapping and contributing towards a combined deeper absorption feature. This will either result in a failed convergence with the number of steps exceeding $20 \tau$, or the process can converge, but with a set of parameters that can be seen to poorly fit the multiple components in the observations. 

It is also possible for SB2 candidates to show up in another way. This is exemplified by two particular targets, SSN~13 and SSN~15. These  targets have strong \ion{He}{ii} lines, and in both cases the Bayesian statistic technique finds an RV fit that shifts the observations to closely match these lines. The weak \ion{He}{i} lines are then shown to be best fit by a different RV value. This, in combination with the RV shifts evident for both targets, leads us to categorise both targets as SB2 candidates, where the \ion{He}{ii} lines are the result of a hot primary and the \ion{He}{i} lines are due to a cooler secondary. SB2 and SB3 candidates identified through any means are noted separately, and the best-fit parameters are not included in the results.

%________________________________________________________________
\section{Results}
\label{sec:results}

The stellar parameters results ($T_\ast$, $\log g$, and $\varv \sin i$) are   from the sampler return of the MCMC process. With the burn-in samples discarded, the median result serves as the best-fit result for these parameters. Table~\ref{tab:NGC346_MUSE_results} lists the  single-star fit found for 34 stars. These results include the error estimations for each stellar parameter (Sect.~\ref{sec:error_estimation}).
Histograms of the measured temperature, surface gravity, projected rotational velocity, luminosity, and spectroscopic mass of the targets within the sample are shown in Fig.~\ref{fig:stellar_params_hist}. They show that the majority of the targets have $T_\ast > 30$\,kK, showing the reliance on prominent \ion{He}{}spectral lines. The distribution shows far fewer targets with $T_\ast > 34 \mathrm{kK}$.
Among the stars in our sample, ten stars meet these criteria and are designated accordingly as SB1.

% %________________________________________________________________

\section{Discussion}
\label{sec:discussion}

\subsection{Rejected results for individual stars}
\label{sec:failure_to_converge}
Over 200 targets have at least one observation meeting the minimum S/N criteria, yet only 34 have  single-star results where the result is accepted (Sect.~\ref{sec:results}). Whether or not the process produces a result for a target that is accepted based on inspection is not influenced by the number of observations available, with some of the accepted results being for objects with as few as two observations. Even with multiple high S/N observations, a target without \ion{He}{ii} lines cannot be fit. Without the balance of the \ion{He}{i} and \ion{He}{ii} lines, the temperature becomes less constrained. As the parameters of $T_\ast$, $\log g$, and $\varv \sin i$ are partially degenerate when fitting using only the H and He diagnostic lines, 
the lack of information about temperature from the balance of \ion{He}{i} and \ion{He}{ii} lines results in too little information to constrain all three parameters simultaneously.
These features of the Bayesian statistic technique give it a selection bias. Lower temperature objects and peculiar objects are necessarily excluded from the final results.

%---------------------------------------------------------------
\begin{figure}[tbp]
        \centering
        \includegraphics[width=\hsize]{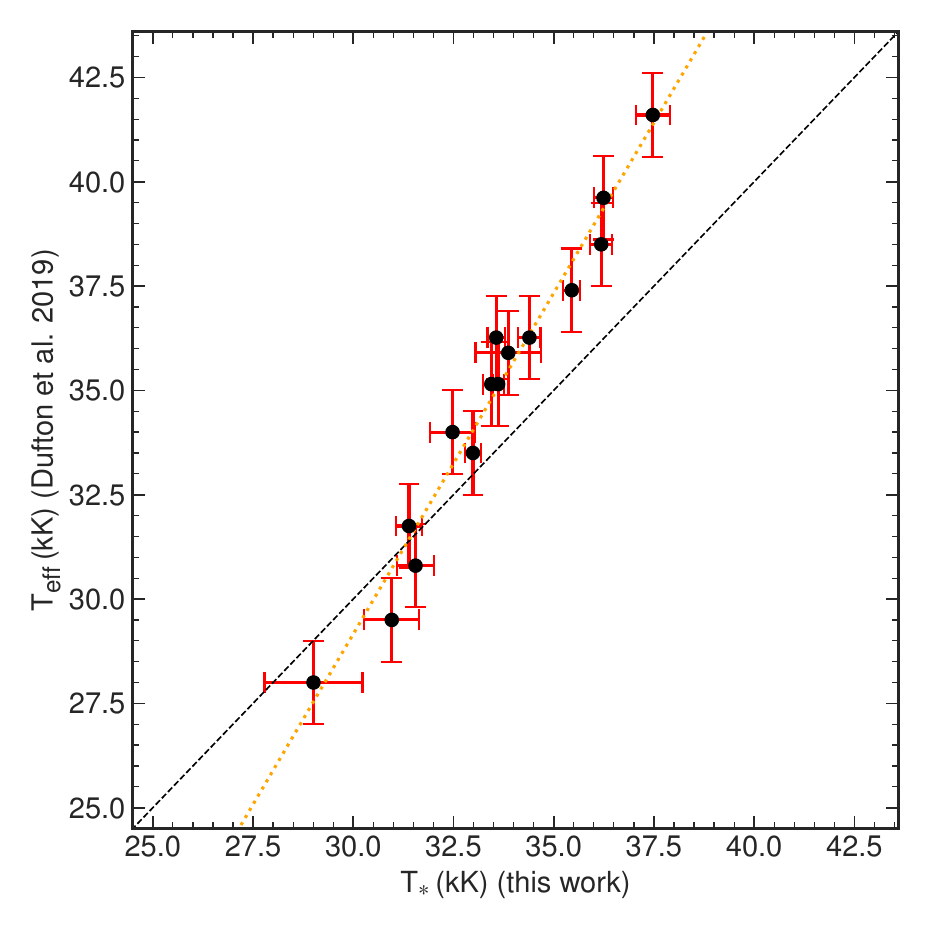}
        \caption{
                Comparison of fitted temperature   found via the Bayesian statistic technique used in this work to that found by \citet{Dufton+2019} fitting to \textsc{tlusty} grid models. The stars shown are limited to those in the overlap between the two samples. The black dashed line shows a gradient of one, where the two methods would have complete agreement. The orange dashed line shows the best-fit line between $T_\ast$ of the two samples.
        }
        \label{fig:Teff_comparison}
\end{figure}
%---------------------------------------------------------------

%---------------------------------------------------------------
\begin{figure}[tbp]
        \centering
        \includegraphics[width=\hsize]{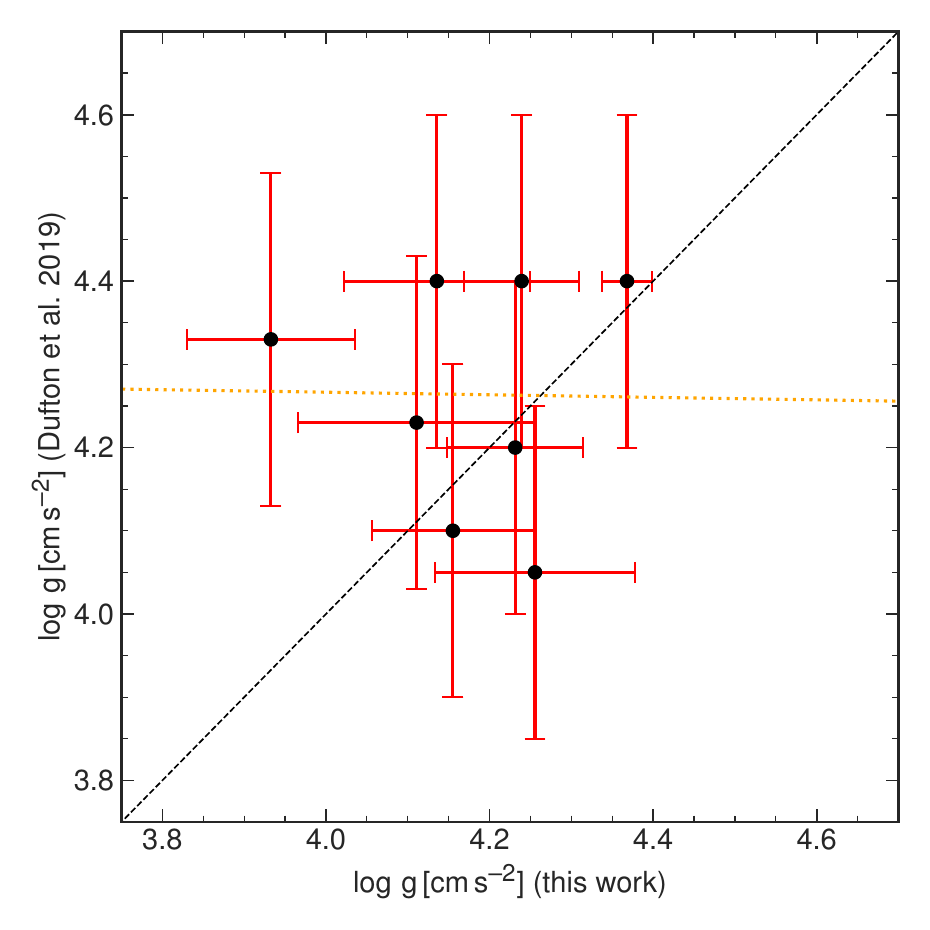}
        \caption{
                Comparison of surface gravity   found via the Bayesian statistic technique used in this work to that found by \citet{Dufton+2019} fitting to \textsc{tlusty} grid models. The stars shown are limited to those in the overlap between the two samples. The dashed lines are coloured as in Fig.~\ref{fig:Teff_comparison}.
        }
        \label{fig:logg_comparison}
\end{figure}
%------------------------------------------------------

%---------------------------------------------------------------
\begin{figure*}
        \centering
        \includegraphics[width=\hsize]{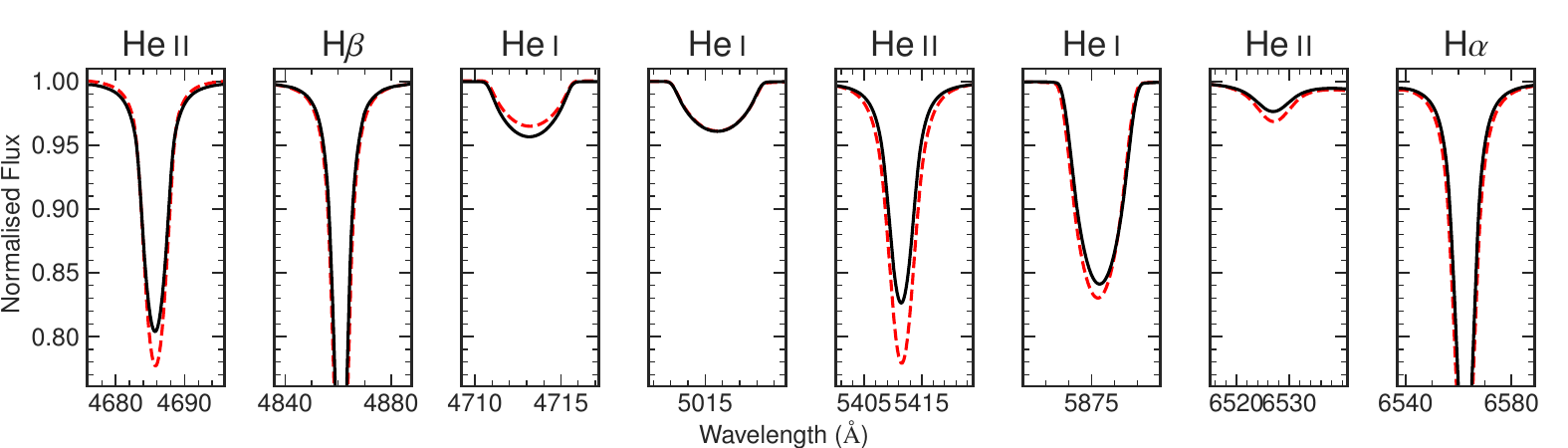}
        \caption{
                Synthetic spectral lines created from PoWR (red dashed line) and \textsc{tlusty} (black solid line) grid models. Each synthetic spectrum is for a model with the same temperature ($T_\ast = 40 \mathrm{kK}$) and surface gravity ($\log g = 4.0$). Each synthetic spectra has been convolved with a rotation profile for $\varv \sin i = 150 \mathrm{km\,s}^{-1}$.
        }
        \label{fig:tlusty_model_comparisons}
\end{figure*}
%--------------------

\subsection{Comparison to previous works}
\label{sec:comp_fit_method}

The population of massive stars in NGC\,346 have been catalogued before \citep{Dufton+2019}, creating the opportunity to compare the results in this work with other methods for estimating stellar parameters. Sixteen of the 34 stars in the sample of single stars have been previously studied (see \citealt{Dufton+2019}). 
They find $T_\mathrm{eff}$ and $\log g$  by fitting the \textsc{flames} spectra to synthetic spectra generated from \textsc{tlusty} and \textsc{synspec} model codes \citep{Hubeny1988, HubenyLanz1995, Hubeny+1998, LanzHubeny2007}. This code produces publicly available \textsc{tlusty} grids of synthetic spectra.\footnote{\url{http://tlusty.oca.eu/}} The estimated errors in \citet{Dufton+2019} are $1 \,\mathrm{kK}$ in $T_\mathrm{eff}$ and 0.2~dex in $\log g$. Figures~\ref{fig:Teff_comparison} shows a comparison between the fitted temperature values and Fig.~\ref{fig:logg_comparison}  shows a comparison of  the $\log g$ values.

Figure~\ref{fig:Teff_comparison} shows that there is no  agreement between the temperature results of the overlapping samples, but that both results agree as to the temperature order of the sample (i.e. hotter stars are identified as hotter objects via both methods). However, there is a difference, especially at higher temperatures: the method based on the PoWR grid of synthetic spectra generally does not find  temperatures as high as those found by \citet{Dufton+2019} via matching to the \textsc{tlusty} grid of synthetic spectra. This difference warrants further investigation.

Some of this difference will come down to the choice of line selection used by \citet{Dufton+2019} to find the best-fitting \textsc{tlusty} synthetic spectrum, such as the use of \HeI{4922} and \HeII{4686}. These lines are not included for the method used in this work, due to the reasons detailed in Sect.~\ref{sec:line_segment_selection}. Further differences can be explained by the difference between the PoWR models used to generate the PoWR grid of the synthetic spectra and the \textsc{tlusty} and \textsc{synspec} model codes used for the \textsc{tlusty} grid of the synthetic spectra. 

As an example, Figure~\ref{fig:tlusty_model_comparisons} shows the H$\alpha$, H$\beta$, \ion{He}{i}, and \ion{He}{ii} line segments from a synthetic spectrum taken from the \textsc{tlusty} grid for $T_\ast = 40 \mathrm{kK}$ and $\log g = 4.0$, convolved with a rotation profile equivalent to $\varv \sin i = 150 \mathrm{km\,s}^{-1}$. This is shown in comparison to the same line segments from the PoWR grid synthetic spectrum for the same stellar parameters. It can be seen immediately that the PoWR synthetic spectrum shows stronger \ion{He}{ii} lines. This means that when using this PoWR grid, a star of this temperature will be fitted to a synthetic spectrum generated with a lower temperature PoWR model than if using a grid of synthetic spectra generated with the \textsc{tlusty} and \textsc{synspec} model codes.

A detailed investigation into the differences between the PoWR model code and the \textsc{tlusty} model code is beyond the scope of this work, but a brief set of reasons why the two codes may result in such a difference in the resultant synthetic spectra may include the following: the difference in metal abundances adopted, the difference between the selected mass loss recipe used, the difference in the treatment of line blanketing, the differences in implementation of microtubulence as an additional broadening mechanism, and the luminosity selected for each model. With such differences in the  set-up of the models, it is not surprising that the overlapping sample shows different stellar parameter results between the two samples. By way of reassurance, the line comparison figure for each target is included on Zenodo, and an example is shown in Fig.~\ref{fig:SSN22_lines_comparisons_plot}. These show how the observations are each well fit by a synthetic spectrum based on the grid of spectra generated by the PoWR model code.

The MUSE observations included in this work were also employed in the analysis presented in \citet{Rickard+2022}. This previous work focused on the wind parameters of massive stars in the central core of NGC\,346 using PoWR models. This required PoWR models to be run to determine the wind parameters. These models use the input of stellar parameters determined from optical spectra. For the majority of the brightest objects within the sample, previous studies such as \citet{Bouret+2003} and \citet{Dufton+2019} provided the stellar parameters of $T_\ast$, $\log g$ and $\varv \sin i$. For the remainder, where no previous analysis existed, a mean MUSE spectrum was generated from each target, with each epoch shifted by a suitable RV found through a shift-and-add process. For these, $\varv \sin i$ was measured using the \textsc{iacob-broad} tool from He lines. The best-fitting synthetic spectrum from the grid generated from PoWR models was then selected using a $\chi^2$ test, with changes to $T_\ast$ and $\log g$ applied to the custom PoWR models only in the small number of cases where it was clearly needed.

Despite the difference in method, the results found in this work are very similar to those results presented in \citet{Rickard+2022}. For the eight objects where \citet{Rickard+2022} uses the MUSE spectra to determine stellar parameters, the average difference between the results presented there and those found here with the Bayesian statistic technique is $0.7 \mathrm{kK}$ for $T_\ast$ and $0.1 \mathrm{dex}$ for $\log g$, well within the steps in the PoWR grid of $1 \mathrm{kK}$ for and $0.2 \mathrm{dex}$
This suggests that the difference between these results and those found by \citet{Dufton+2019} are due to the underlying differences in the model codes and the model set-up choices and not are not the result of the Bayesian statistic technique.

Figure~\ref{fig:logg_comparison} shows the derived $\log g$ in this work compared to that found in \citet{Dufton+2019}. There is much greater scatter for this parameter and a less clear trend, due to the difficulty in measuring $\log g$ from the wings of the Balmer lines via  both methods. In addition, $T_\ast$ and $\log g$ are partially degenerate and measured simultaneously. If the two methods do not find the same $T_\ast$, they will not agree in their measurement of $\log g$. Regardless, there is agreement for both of the overlapping objects, with both methods agreeing on which targets show indications of lower $\log g$ and which show indications of higher $\log g$.

%---------------------------------------------------------------
\begin{figure}[tbp]
        \centering
        \includegraphics[width=\hsize]{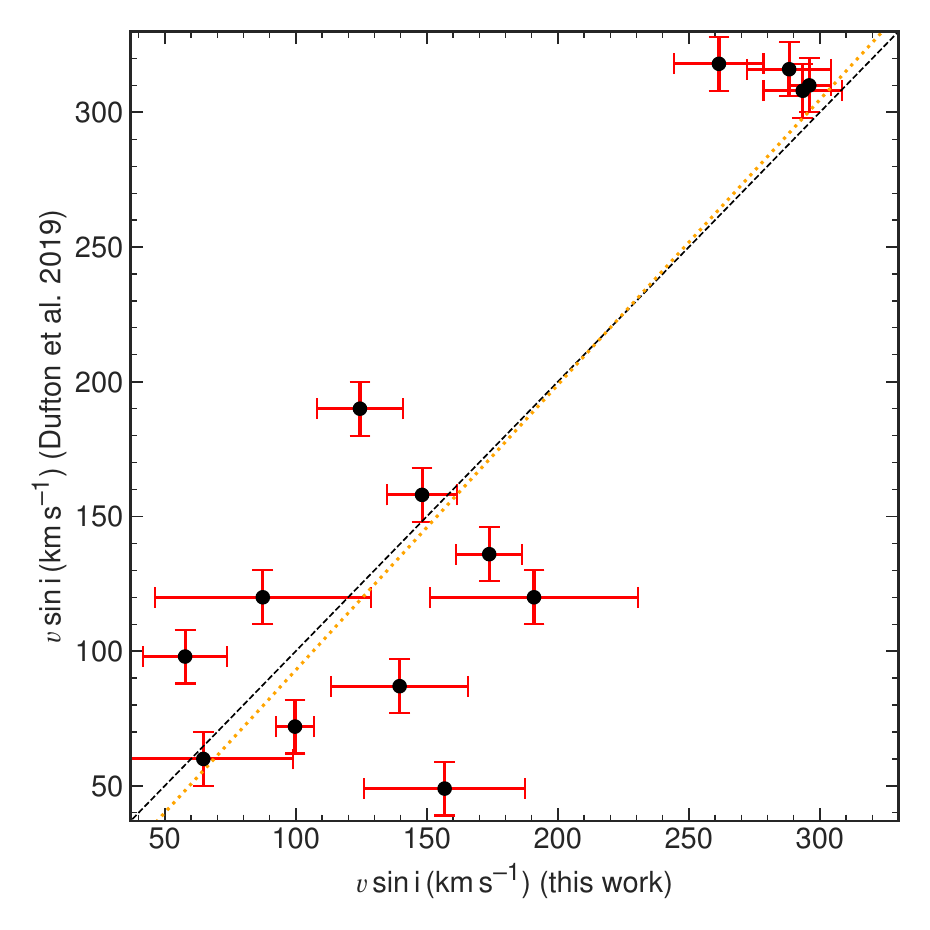}
        \caption{
                Comparison of the projected rotational velocity as found from the Bayesian statistic technique used in this work on  H and  He lines along with that found by \citet{Dufton+2019} using the goodness of fit method to \ion{Mg}{} and \ion{Si}{}. Stars shown are limited to those in the overlap between the two samples. The dashed line colours are the same as in  Fig.~\ref{fig:Teff_comparison}.
        }
        \label{fig:vrot_comparison}
\end{figure}
%---------------------------------------------------------------

In \citet{Dufton+2019} $\varv \sin i$ is measured using Mg and Si lines with a goodness of fit (GF) method and a Fourier transform (FT) method  \citep[\textsc{iacob-broad},][]{Simon-DiazHerror2014}. These lines are outside  the MUSE wavelength range. Figure~\ref{fig:vrot_comparison} shows the overlapping samples and the comparison between the GF results from \citet{Dufton+2019} and to the fitted $\varv \sin i$  in this work. By comparing the GF method results from \citet{Dufton+2019} we are, for both cases, considering a method that not only includes rotation, but also macroturbulence. From the difference between the GF and FT results presented in \citet{Dufton+2019} of $\approx 11 \mathrm{km\,s}^{-1}$, we can judge that the overestimation of our $\varv \sin i$ value due to the inclusion of all broadening affects at once in to one parameter to be of a similar scale of $\approx 11 \mathrm{km\,s}^{-1}$.

It can be seen in Figure~\ref{fig:vrot_comparison} that overall the agreement trend for $\varv \sin i$ in the overlapping sample is very good. However this is a false impression generated by the four objects with rotation $\gtrsim 250 \mathrm{km\,s}^{-1}$. It is clear that both methods correctly identify objects with indications of high broadening velocity in their spectra. In the overlapping sample with rotational broadening velocity $\lesssim 200 \mathrm{km\,s}^{-1}$ there is more disagreement.

While \citet{Zeidler+2022} used the same MUSE observations to derive RVs of 103 stars within the core of NGC{\,}346 for the purpose of understanding the internal kinematics, and this sample would provide a useful verification of the technique used for measuring RV, no overlap exists between the 103 stars reported and this sample  because for most of the stars in their sample, \citet{Zeidler+2022} use metal lines such as \ion{Mg}{i} and \ion{Ca}{ii}, which are not present in the spectra of the hot stars fitted in this work, or they use very strong \ion{He}{i} lines, which would be less prominent in these hot stars with strong \ion{He}{ii}.

\subsection{Binary candidates and binary fraction}

The requirements for a target to be considered SB1 are stringent (Sect.~\ref{sec:sb1}), yet this method finds additional SB1s with respect to those found by \citet{Dufton+2019}. A total of nine SB1 candidates are identified here from the RV shifts from 11 days of MUSE observations.

The selection criteria for \citet{Dufton+2019} was for there to be an RV shift of $> 10\,\mathrm{km\,s}^{-1}$ between two epochs. They identify two SB1 candidates that are not found to be SB1s based on these MUSE observations. These are SSN~33 and SSN~34. 
A review of the violin plot for SSN~33 shows that the variation does not meet the $> 10\,\mathrm{km\,s}^{-1}$ requirement. For SSN~34, the RV distribution shown in the violin plot does exceed the $> 10\,\mathrm{km\,s}^{-1}$ requirement, and shows a clear trend over the 11 days of the observations, but the $4\sigma$ significance test is not satisfied.

The targets of SSN~13 and SSN~15 are considered SB2 candidates, due to the difference in the RV shift required to fit the observed \ion{He}{i} lines compared to the RV shift required to fit the observed \ion{He}{ii} lines. \citet{Dufton+2019} categorise SSN~13 as an SB1 candidate, while SSN~15 matches the criteria used in this work to qualify as an SB1, adding to the evidence of binarity for these targets. An additional target (SSN~47) identified as an SB1 candidate by \citet{Dufton+2019}  has been considered here as an SB2 candidate, due to clear double-line features in observations from a number of epochs.

\begin{table}
\footnotesize
\center
\caption{Binary candidates or emission stars}
\begin{tabular}{ c c } 
\hline
\hline
\centering
Type &  Identification  \\
\hline
\multirow{2}{*}{SB2} & SSN~7, SSN~13, SSN~15, SSN~17, SSN~39, \\
& SSN~47, SSN~58, SSN~80, SSN~89 \\[.5mm]
SB3 & SSN~11 \\[1.5mm]
Be &  SSN~60, SSN~73\\[1mm]

\hline     
\end{tabular}
 % \tablefoot{Stars withing our sample which were not fitted because of composite spectra or strong emission lines in their spectra}
\label{table:misc_stars}

\end{table}

We replicate the SB2 candidate classification from \citet{Dufton+2019} for SSN~39 and see the SB2 features already documented in the observations of SSN~7 \citep{RickardPauli2023}. In addition, we categorise SSN~17, SSN~58, SSN~80, and SSN~89 as SB2 candidates. We believe we can see three distinct line features in the observations of SSN~11, making it an SB3 candidate. A complete list of targets considered but not fit due to binary features within the target's spectrum indicating more than one component, or due to spectral emission features, are listed in Table~\ref{table:misc_stars}. The best-fit result for SSN~13 and SSN~15 is presented in the additional information available on Zenodo\footnote{\url{10.5281/zenodo.13991997}} and show the best-fit model, demonstrating how the \ion{He}{i} and \ion{He}{ii} lines require differing RV shifts. The observations of the remaining objects in Table~\ref{table:misc_stars} are included in this additional information.

There are a total of 34 stars for which we have presented a single-star result. This includes both apparent single stars and SB1s. We do not report the undoubtedly erroneous fit results for SB2 and SB3 candidates and emission stars. With these stars included, the true total number of objects reviewed in this work is 46. If we accept all the SB1 classifications from \citet{Dufton+2019} and consider them in combination with the classifications from this work, we find 11 SB1 candidates, 9 SB2 candidates, and 1 SB3 candidate, resulting in a lower limit of the binary fraction of $21 / 46 = 46\%$. It is worth noting that our method based on the MCMC is only suitable for objects displaying \ion{He}{ii} lines in their spectra, and therefore the derived binary fraction probes only the hottest stars in the central core of NGC\,346.

%---------------------------------------------------------------
\begin{figure}
        \centering
        \includegraphics[width=\hsize]{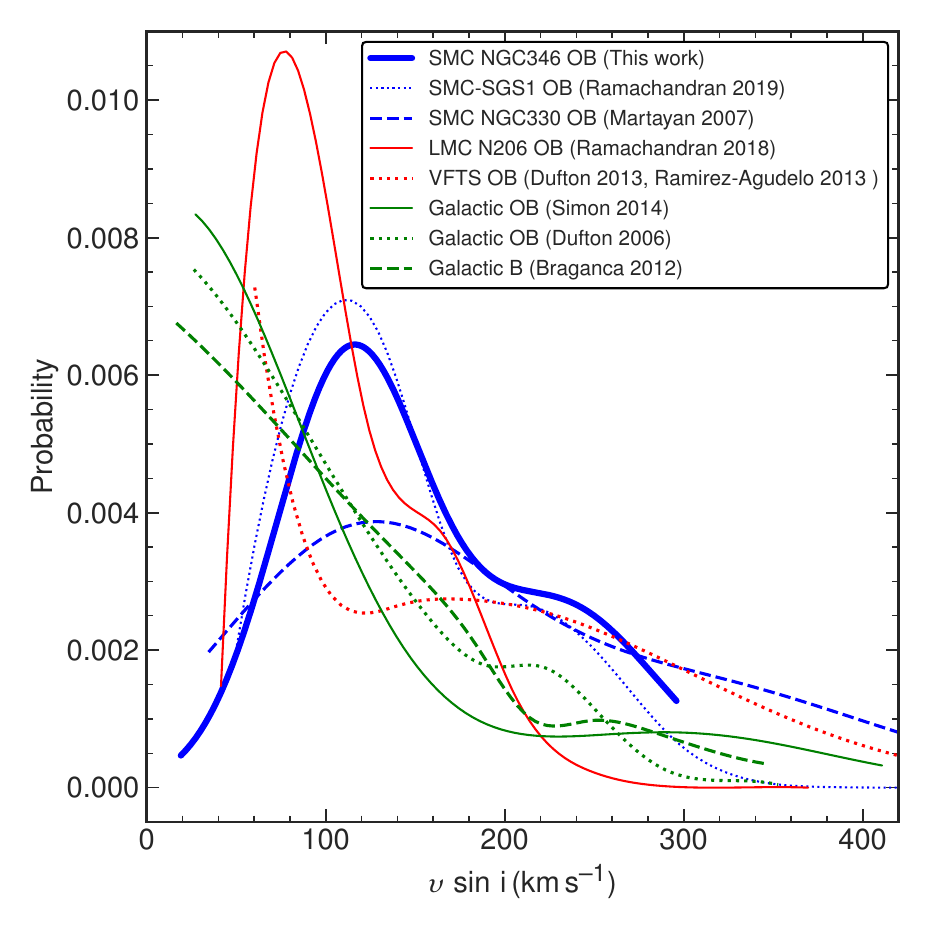}
        \caption{
                Projected rotational velocity probability distribution compared to numerous other SMC, LMC, and Milky Way samples. \emph{Blue lines}: SMC samples--\emph{Blue solid line}: this work (including macroturbulence, and thus a slight overestimation of $\sim 10 \mathrm{km\,s}^{-1}$). \emph{Blue dotted line}: \citet{Ramachandran+2019}. \emph{Blue dashed line}: \citet{Martayan+2007}. \emph{Red lines}: LMC samples--\emph{Red solid line}: \citet{Ramachandran+2018}. \emph{Red dotted line}:  \citet{Dufton+2013} and \citet{Ramirez-Agudelo+2013}. \emph{Green lines}: Galactic samples--\emph{Green solid line}: \citet{Simon-DiazHerror2014}. \emph{Green dashed line}:  \citet{Braganca+2012}. \emph{Green dotted line}:  \citet{Dufton+2006}.
        }
        \label{fig:vsini_pdf}
\end{figure}
%---------------------------------------------------------------

%---------------------------------------------------------------
\begin{figure}
        \centering
        \includegraphics[width=0.99\hsize]{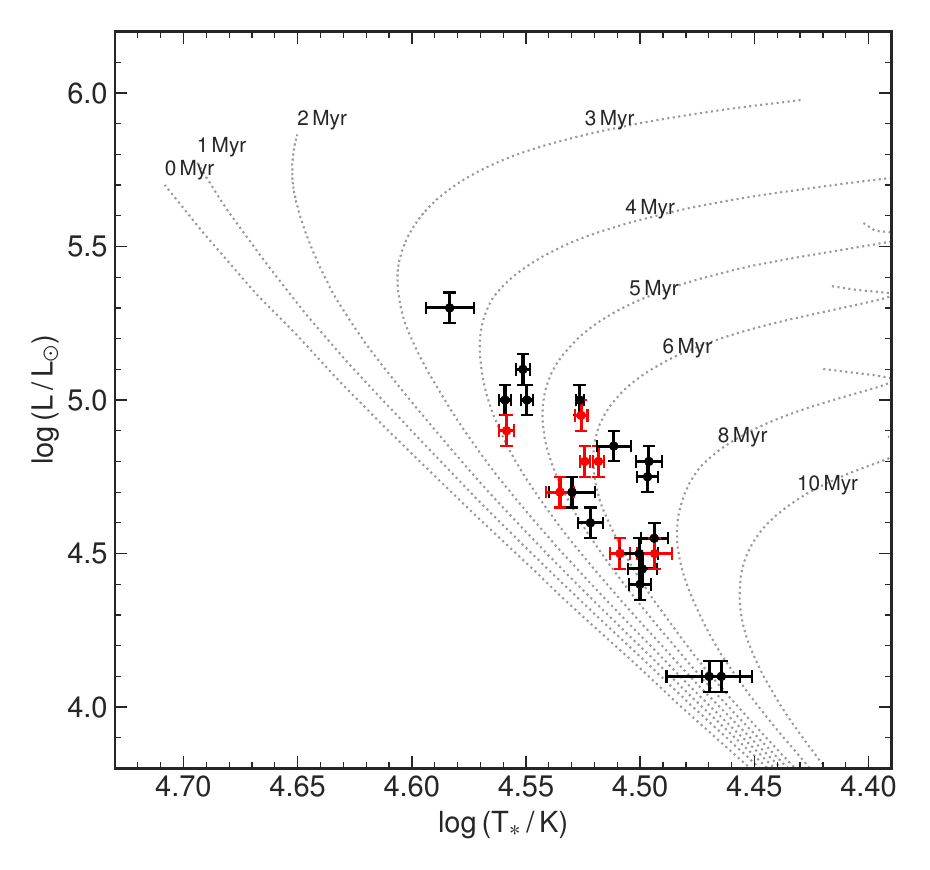}
        \caption{
                HRD of OB stars in the core of NGC{\,}346. \emph{Red markers}: SB1s candidates. \emph{Black markers}: Apparent single stars. \emph{Dotted lines}: Isochrones with $Z=0.002$ and initial rotation velocity of $200\,\mathrm{km\,s}^{-1}$\citep{Georgy+2013}.
        }
        \label{fig:HRD}
\end{figure}
%---------------------------------------------------------------

%--------------------

\subsection{Rotation}

When measuring the projected rotational broadening of a massive star from a single transition line, the lower limit of the rotational velocity that can be measured is limited to the instrument resolution. For this work multiple lines are considered at once, and it is best compared to studies matching the whole spectrum simultaneously to a convolved synthetic spectrum \citep[e.g.][]{Kamann+2018}. By measuring the model of a rotational broadening effect against multiple lines at once, the limited effect of instrumental broadening is eliminated as the shape of the line profile will be measured multiple times, meaning the sampling by the bins along the wavelength dispersal direction of the sensor is improved.

Massive stars in low-metallicity environments typically have  higher rotational velocities
compared to their Galactic counterparts because their weaker stellar winds remove less mass and angular momentum \citep[e.g.][]{Ramachandran+2019}. 
Figure~\ref{fig:vsini_pdf} shows the probability distribution of rotational velocities in our sample stars in comparison to other SMC, LMC, and Galactic samples. As in previous studies, the sample presented in this work shows the peak in the projected rotational velocity distribution at higher velocities than for the LMC samples, which in turn peaks at higher velocities than the Galactic  samples. This supports the argument that lower metallicity massive stars spin down more slowly than their higher metallicity equivalents, due to their lower stellar wings and a smaller resultant loss of angular momentum. However, caution must be taken
when comparing different samples. For example, the OB star sample presented by \citet{Ramachandran+2019} includes later-type stars which are not included in our sample, due to the lack of strong \ion{He}{ii} lines. In addition, the $\varv \sin i$ values found in this work include additional broadening mechanisms such as macroturbulence, and there is a slight overestimation of $\sim 10\,\mathrm{km\,s}^{-1}$

Despite this, it is interesting to note that the location of the peak of the probability distribution  of rotation rate in our sample is very similar
to that found in previous studies of SMC objects, such as OB stars in the SMC Wing \citep{Ramachandran+2019} and in the NGC~330 cluster  \citep{Martayan+2007}. This  can be explained by the lack of a correlation between stellar temperature and the projected rotation rate, meaning that the absence of cooler B stars in our sample does not affect the projected rotational velocity distribution.  

% %________________________________________________________________

\subsection{Estimating  the age of the NGC\,346 star cluster}
\label{sec:stellar_evo}

One method to estimate the age of a stellar cluster is to compare the Hertzsprung--Russell diagram (HRD) position and main sequence turnoff to sets of model isochrones with the modelled ages. The best-fit values of luminosity and temperature  are plotted on a  HRD in Fig.~\ref{fig:HRD}. This only includes the stars with available HST UV observations, which we used to derive luminosities (Sect.~\ref{sec:lum_fit}). The  HRD positions of our sample stars are compared to isochrones with an initial rotation of $200\,\mathrm{km\,s}^{-1}$ (i.e. matching the projected rotational velocity found in our sample stars). As can be seen in Fig.~\ref{fig:HRD}, the positions of hot OB stars in our sample are  consistent with a minimum age of at least $3\,\mathrm{Myr}$. 

This is higher than the previous age estimates of $1{-}2.6$~Myr \citep{Dufton+2019} and $1{-}2$~Myr \citep{Walborn+2000}. The reason for this discrepancy is that the previous works relied on the HRD position of the hotter stars, such as SSN~7 and SSN~9. Care should be taken when using these stars for the age estimates. The very bright star SSN~9 is a giant  with high nitrogen content 
\citep{Walborn+2000, Bouret+2013, Rickard+2022}. 
The other bright star, SSN~7, is a  SB2 with two high-mass components that have already exchanged mass \citep{RickardPauli2023}. The evolutionary age of the   SSN~7 system is $4.2\,\mathrm{Myr}$ \citep{RickardPauli2023}. Therefore, treating it as a single star and using its 
location on the HRD for comparison with isochrones for cluster age estimates is misleading.  The HRD position of the OB stars in the sample of stars we consider in this work provide further support for the estimated  minimum age of  OB stars in the core of NGC{\,}346 as $\gtrsim 3\,\mathrm{Myr}$ \citep{Rickard+2022, RickardPauli2023}.

\section{Summary and conclusions}
\label{sec:conclusions}

This work derives the fundamental stellar parameters, $T_\ast$, $\log g$, and $\varv \sin i$ for a sample of 34 OB-type stars located in the core of the NGC{\,}346 cluster in the SMC. Multi-epoch spectra were obtained with the MUSE IFU spectrograph. The spectroscopic analysis presented in this paper is based on a Bayesian statistic technique. This method is independent of a measurement of projected rotational velocity made through other means and breaks the degeneracy between these three parameters when only considering \ion{H}{}and \ion{He}{}spectral features. The results are compared to a subsample of 18 stars where these parameters were found using standard methods.

We conclude that the Bayesian statistic technique is an effective method for deriving the most likely stellar parameters of hot OB-type stars using optical spectroscopy and pre-calculated grids of model synthetic spectra. We note that this method for determining temperature, surface gravity, and projected rotational velocity does not require metal lines, and thus can be used for studies of metal-poor massive stars. In order to break the partial degeneracy between the three stellar parameters, clear \ion{He}{ii} lines are required, which means that this method can only be applied to hotter stars (i.e. mainly O stars).

In a subsample of stars in the overlap between this work and the sample of OB stars   analysed by \citet{Dufton+2019}, there is agreement as to relative temperature through the course of the samples, but there is disagreement regarding the best-fit values for $T_\ast$, which consequently affects the best-fit value for $\log g$ and $\varv \sin i$. However, this difference is due to the difference between the model code used to generate the grid of synthetic spectra employed in this work (PoWR) and the model code used to generate the grid of synthetic spectra employed by \citet{Dufton+2019} (\textsc{tlusty} and \textsc{synspec}).

The peak in the probability distribution of the projected rotational velocity is at a higher velocity than that of the Galactic and LMC OB populations, in support of other findings that SMC stars have higher rotation. The peak is at a  projected rotational velocity similar to  that found for other clusters of SMC stars.

Combining multiple epochs allows   the detection of binaries. Combining this method with the binary results from other work we obtain a minimum binary fraction of the hot OB stars from the core of NGC{\,}346 of $> 46\%$.

The use of synthetic SEDs for each target compared to available HST UV spectra and photometry allowed   the fitting of stellar luminosity. The resultant HRD position of each target supports the previously identified minimum age of the OB population of the core of NGC~346 to be $\gtrsim 3\,\mathrm{Myr}$.

%%%%%%%%%%%%%%%%%%%%%%%%%% ONLINE MATERIAL %%%%%%%%%%%%%%%%%%%%%%%%%%%%%%%%%%%%
\section*{Data availability}
\label{onlinematerial}
% \Online
The discussion and results from each target included in this work and the observations of the SB2 and SB3 candidates and emission stars can be accessed on the Zenodo repository (\url{10.5281/zenodo.13991997}).

%__________________________________________________________________
\begin{acknowledgements}

We thank the anonymous referee for their constructive comments.
        DP acknowledges financial support by the Deutsches Zentrum  f\"ur  Luft  und  Raumfahrt  (DLR)  grant  FKZ  50  OR  2005.
        This publication has benefited from a discussion at a team meeting sponsored by the International Space Science Institute at Bern, Switzerland. 
        L.\,M.\,O. acknowledges support from the Verbundforschung grant 50 OR 1809.
Some/all of the data presented in this paper were obtained from the Mikulski Archive for Space Telescopes (MAST). STScI is operated by the Association of Universities for Research in Astronomy, Inc., under NASA contract NAS5-26555. Support for MAST for non-\textsl{HST} data is provided by the NASA Office of Space Science via grant NNX09AF08G and by other grants and contracts.
AACS and VR acknowledge support by the Deutsche Forschungsgemeinschaft (DFG, German Research Foundation) in the form of an Emmy Noether Research Group -- Project-ID 445674056 (SA4064/1-1, PI Sander). 
ECS acknowledges financial support by the Federal Ministry for Economic Affairs and Climate Action (BMWK) via the German Aerospace Center (Deutsches Zentrum f\"ur Luft- und Raumfahrt, DLR) grant 50 OR 2306 (PI: Ramachandran/Sander).
AACS, VR, and ECS further acknowledge support from the Federal Ministry of Education and Research (BMBF) and the Baden-Württemberg Ministry of Science as part of the Excellence Strategy of the German Federal and State Governments.
This research has made use of the VizieR catalogue access tool, Strasbourg, France. The original description of the VizieR service was published in A\&AS 143, 23.
Part of work was facilitated by the International Space Science Institute (ISSI) in Bern, through ISSI International Team project 512 (Multiwavelength View on Massive Stars in the Era of Multimessenger Astronomy, PI Oskinova).

\end{acknowledgements}

\bibliographystyle{aa}
\bibliography{paper}

\begin{appendix} %first appendix
\onecolumn
\smallskip\noindent
\section{Additional tables}
\label{app:additional_tables}

The full list of MUSE observations, including observing conditions, is included in Table~\ref{table:obs_cond}. The full results for each target is included in Table~\ref{tab:NGC346_MUSE_results}.

%%%%%%%%%%%%%%%%%%%%%%%%%%%%%%%%%%%%%%%%%%%%%%%%%%%%%%%%%%%%%%%%%%%%
%%%%%%%% MUSE: observing conditions
%%%%%%%%%%%%%%%%%%%%%%%%%%%%%%%%%%%%%%%%%%%%%%%%%%%%%%%%%%%%%%%%%%%%
\begin{table*}[!h]
    % \tiny
    \caption{Observation diary of the individual MUSE observation blocks.}
        \label{table:obs_cond}
        \centering  
        \begin{tabular}{cccccc}
                \hline\hline %---------------------------------------
                Observation date & Observation Mid Time & MJD & Observation Duration & Seeing & Airmass \\
                (DD/MM/YYYY) & (UTC) &  & (s) & (\arcsec) &     \\
                \hline  %---------------------------------------------------------------------
        11/08/2016 &    07:54:12        &  57611.32930492  & $8\times315$ & 1.94  &  1.5  \\
        13/08/2016      & 05:54:40 &  57613.24629853  &  $8\times315$ &1.25  &  1.6  \\
        13/08/2016      & 06:55:08      &  57613.28828952  & $8\times315$ & 1.18  &  1.5  \\
        17/08/2016 &    05:37:52 &  57617.23462962  & $8\times315$ & 1.17  &  1.6  \\
        18/08/2016      & 02:28:43      &  57618.10327055  & $8\times315$ & 1.81  &  2.3  \\
        20/08/2016      & 03:34:04      &  57620.14865186  & $8\times315$ & 1.84  &  1.9  \\
        20/08/2016      & 04:34:12      &  57620.19041780  & $8\times315$ & 1.63  &  1.7  \\
        20/08/2016      & 05:34:58      &  57620.23261681  &$8\times315$ &  1.53  &  1.6  \\
        21/08/2016      & 04:57:13      &  57621.20640006  & $8\times315$ & 1.44  &  1.6  \\
        21/08/2016      & 05:57:37      &  57621.24834448  &$8\times315$ &1.56  &  1.5  \\
        22/08/2016      & 03:31:57      &  57622.14718256  & $8\times315$ & 1.42  &  1.9  \\
                \hline %---------------------------------------------------------------------
        \end{tabular}
\end{table*}
%%%%%%%%%%%%%%%%%%%%%%%%%%%%%%%%%%%%%%%%%%%%%%%%%%%%%%%%%%%%%%%%%%%%
%%%%%%%%%%%%%%%%%%%%%%%%%%%%%%%%%%%%%%%%%%%%%%%%%%%%%%%%%%%%%%%%%%%%

% \vspace{.5cm}

\begin{table*} 
    \centering 
    \tiny 
    \caption{MCMC results of sample of stars within NGC 346.} 
    \begin{tabular}{cccccccccc} \hline \hline \rule{0cm}{2.2ex} 
    SSN ID & \multicolumn{2}{c}{Spectral Type} & $T_\ast$ & $\log g$ & $\varv \sin i$ & $\log L_\ast$ & E(B{-}V) & Spec Mass\textsuperscript{$\ddagger$} & Note \\ 
    & \citep{Massey+1989} & (from $T_\ast$\textsuperscript{$\dagger$}) & (kK) & [$\mathrm{cm\,s}^{-2}$] & ($\mathrm{km \, s}^{-1}$) & [$L_\odot$] & & ($M_\odot$) \\[1mm]
    \hline 
   18 &    O6.5V &    O7.5V &     $35.6\pm 0.2$ &     $4.21\pm 0.06$ &     $130\pm 11$ &     $5.10 \pm 0.05 $ &     $0.12 \pm 0.01 $ &     $51.6 \pm 7.3$ &     ... \\[1mm]
   22 &    ... &    O6.5V &     $38.3\pm 0.9$ &     $3.88\pm 0.16$ &     $110\pm 20$ &     $5.30 \pm 0.05 $ &     $0.12 \pm 0.01 $ &     $28.2 \pm 10.1$ &     ... \\[1mm]
   31 &    ... &    O7.5V &     $36.2\pm 0.2$ &     $4.14\pm 0.04$ &     $58\pm 16$ &     $5.00 \pm 0.05 $ &     $0.13 \pm 0.01 $ &     $32.5 \pm 2.6$ &     ... \\[1mm]
   32 &    B0V &    O8.5V &     $33.0\pm 0.2$ &     $4.37\pm 0.03$ &     $296\pm 8$ &     $4.80 \pm 0.05 $ &     $0.13 \pm 0.01 $ &     $50.5 \pm 3.5$ &    SB1\\[1mm]
   33 &    O8V &    O7.5V &     $35.5\pm 0.2$ &     $4.23\pm 0.05$ &     $124\pm 16$ &     $5.00 \pm 0.05 $ &     $0.14 \pm 0.01 $ &     $43.4 \pm 4.9$ &     ... \\[1mm]
   34 &    O9.5V &    O8.5V &     $33.6\pm 0.1$ &     $4.19\pm 0.03$ &     $148\pm 13$ &     $5.00 \pm 0.05 $ &     $0.13 \pm 0.01 $ &     $49.3 \pm 3.2$ &     ... \\[1mm]
   36 &    O8V &    O8.5V &     $33.6\pm 0.2$ &     $4.02\pm 0.08$ &     $140\pm 26$ &     $4.95 \pm 0.05 $ &     $0.12 \pm 0.01 $ &     $30.0 \pm 5.4$ &    SB1\\[1mm]
   40 &    B0V &    O9.0V &     $31.3\pm 0.4$ &     $4.09\pm 0.16$ &     $242\pm 25$ &     $4.80 \pm 0.05 $ &     $0.12 \pm 0.01 $ &     $32.5 \pm 11.6$ &     ... \\[1mm]
   41 &    O7.5V &    O8V &     $34.4\pm 0.3$ &     $4.18\pm 0.07$ &     $174\pm 13$ &     ... &    ... &    ... &   SB1\\[1mm]
   43 &    ... &    O8.5V &     $33.4\pm 0.2$ &     $4.28\pm 0.06$ &     $206\pm 21$ &     $4.80 \pm 0.05 $ &     $0.11 \pm 0.01 $ &     $39.0 \pm 5.0$ &    SB1\\[1mm]
   44 &    O8V &    O8.5V &     $33.4\pm 0.2$ &     $4.05\pm 0.04$ &     $100\pm 7$ &     ... &    ... &    ... &    ... \\[1mm]
   46 &    O6V: &    O7.5V &     $36.2\pm 0.3$ &     $4.17\pm 0.07$ &     $288\pm 16$ &     $4.90 \pm 0.05 $ &     $0.14 \pm 0.01 $ &     $27.9 \pm 4.6$ &    SB1\\[1mm]
   50 &    O7V &    O7V &     $37.5\pm 0.4$ &     $3.93\pm 0.10$ &     $104\pm 12$ &     ... &    ... &    ... &    ... \\[1mm]
   55 &    ... &    O9.0V &     $31.4\pm 0.3$ &     $4.24\pm 0.07$ &     $261\pm 17$ &     $4.75 \pm 0.05 $ &     $0.14 \pm 0.01 $ &     $40.8 \pm 6.6$ &     ... \\[1mm]
   56 &    ... &    O9.0V &     $31.8\pm 0.5$ &     $4.08\pm 0.11$ &     $189\pm 11$ &     ... &    ... &    ... &    ... \\[1mm]
   57 &    ... &    O9.0V &     $32.5\pm 0.6$ &     $4.14\pm 0.11$ &     $191\pm 40$ &     $4.85 \pm 0.05 $ &     $0.14 \pm 0.01 $ &     $35.3 \pm 9.2$ &     ... \\[1mm]
   59 &    ... &    O8.5V &     $33.9\pm 0.8$ &     $4.28\pm 0.21$ &     $217\pm 36$ &     $4.70 \pm 0.05 $ &     $0.11 \pm 0.01 $ &     $29.8 \pm 14.6$ &     ... \\[1mm]
   62 &    O9V &    O8V &     $34.3\pm 0.5$ &     $4.07\pm 0.10$ &     $197\pm 13$ &     $4.70 \pm 0.05 $ &     $0.11 \pm 0.01 $ &     $17.2 \pm 3.8$ &    SB1\\[1mm]
   77 &    ... &    O9.5V &     $31.0\pm 0.7$ &     $4.23\pm 0.08$ &     $293\pm 15$ &     ... &    ... &    ... &    ... \\[1mm]
   83 &    ... &    O9.0V &     $32.3\pm 0.3$ &     $3.95\pm 0.10$ &     $88\pm 12$ &     $4.50 \pm 0.05 $ &     $0.11 \pm 0.01 $ &     $10.6 \pm 2.4$ &    SB1\\[1mm]
   84 &    ... &    O9.5V &     $31.2\pm 0.4$ &     $4.21\pm 0.18$ &     $180\pm 81$ &     $4.55 \pm 0.05 $ &     $0.13 \pm 0.01 $ &     $24.6 \pm 10.1$ &     ... \\[1mm]
   88 &    ... &    O8.5V &     $34.0\pm 0.2$ &     $4.11\pm 0.07$ &     $98\pm 14$ &     ... &    ... &    ... &    ... \\[1mm]
   100 &    ... &    O8.5V &     $33.9\pm 0.8$ &     $4.11\pm 0.15$ &     $65\pm 34$ &     ... &    ... &    ... &    ... \\[1mm]
   102 &    ... &    O9.0V &     $31.6\pm 0.5$ &     $4.26\pm 0.12$ &     $87\pm 41$ &     $4.45 \pm 0.05 $ &     $0.13 \pm 0.01 $ &     $20.8 \pm 5.9$ &     ... \\[1mm]
   105 &    ... &    O9.0V &     $31.6\pm 0.4$ &     $4.17\pm 0.09$ &     $117\pm 24$ &     $4.40 \pm 0.05 $ &     $0.12 \pm 0.01 $ &     $15.0 \pm 3.0$ &     ... \\[1mm]
   109 &    ... &    O9.0V &     $31.7\pm 0.6$ &     $4.05\pm 0.08$ &     $293\pm 19$ &     $4.50 \pm 0.05 $ &     $0.13 \pm 0.01 $ &     $14.4 \pm 2.6$ &     ... \\[1mm]
   110 &    ... &    O8.5V &     $33.3\pm 0.4$ &     $4.17\pm 0.14$ &     $19\pm 26$ &     $4.60 \pm 0.05 $ &     $0.17 \pm 0.01 $ &     $19.3 \pm 6.1$ &     ... \\[1mm]
   122 &    ... &    O9.5V &     $31.2\pm 0.6$ &     $4.09\pm 0.14$ &     $128\pm 27$ &     $4.50 \pm 0.05 $ &     $0.12 \pm 0.01 $ &     $16.9 \pm 5.4$ &    SB1\\[1mm]
   142 &    ... &    O9.5V &     $30.8\pm 0.1$ &     $4.06\pm 0.04$ &     $87\pm 7$ &     ... &    ... &    ... &    ... \\[1mm]
   171 &    ... &    B0V &     $29.0\pm 1.2$ &     $4.16\pm 0.10$ &     $157\pm 31$ &     ... &    ... &    ... &    ... \\[1mm]
   177 &    ... &    B0.2V &     $28.9\pm 0.8$ &     $4.15\pm 0.12$ &     $161\pm 32$ &     ... &    ... &    ... &   SB1\\[1mm]
   188 &    ... &    B0V &     $29.5\pm 1.3$ &     $4.12\pm 0.11$ &     $147\pm 46$ &     $4.10 \pm 0.05 $ &     $0.13 \pm 0.01 $ &     $8.8 \pm 2.3$ &     ... \\[1mm]
   193 &    ... &    B0V &     $29.1\pm 0.6$ &     $4.18\pm 0.06$ &     $71\pm 33$ &     $4.10 \pm 0.05 $ &     $0.14 \pm 0.01 $ &     $10.7 \pm 1.4$ &     ... \\[1mm]
   245 &    ... &    B1V &     $26.1\pm 0.9$ &     $4.14\pm 0.06$ &     $110\pm 32$ &     ... &    ... &    ... &    ... \\[1mm]
    \hline \rule{0cm}{2.4ex} 
    \end{tabular} 
    \tablefoot{\tablefoottext{$\dagger$}{Following relation from \cite{Ramachandran+2019}.} \tablefoottext{$\ddagger$}{Calculated from the temperature, surface gravity, and luminosity results via the Stefan--Boltzmann law}} 
    \label{tab:NGC346_MUSE_results} 
    \rule{0cm}{2.8ex} 
\end{table*}

\end{appendix}

\end{document}